\documentclass[useAMS,usenatbib]{mn2e}
\usepackage{url}
\usepackage{graphicx}
\usepackage{amsmath}
\usepackage{color}

\newcommand{\lir}{$L_{\rm IR}$~}
\newcommand{\luv}{$L_{\rm FUV}$~}
\newcommand{\lirns}{$L_{\rm IR}$}
\newcommand{\luvns}{$L_{\rm FUV}$}
\newcommand{\msun}{\textrm{M}_{\odot}}
\newcommand{\lsun}{\textrm{L}_{\odot}}

\title[Dust attenuation and SFR from $z\sim4$ to $z\sim1.5$]{HerMES:
  dust attenuation and star formation activity in UV-selected samples
  from $z\sim4$ to $z\sim1.5^{\dagger}$} \author[S.~Heinis et al.]
{\parbox{\textwidth}{\raggedright S.~Heinis,$^{1,2}$\thanks{E-mail: \texttt{sheinis@astro.umd.edu}}
V.~Buat,$^{1}$
M.~B{\'e}thermin,$^{3,4}$
J.~Bock,$^{5,6}$
D.~Burgarella,$^{1}$
A.~Conley,$^{7}$
A.~Cooray,$^{8,5}$
D.~Farrah,$^{9}$
O.~Ilbert,$^{1}$
G.~Magdis,$^{3}$
G.~Marsden,$^{10}$
S.J.~Oliver,$^{11}$
D.~Rigopoulou,$^{12,13}$
Y.~Roehlly,$^{1}$
B.~Schulz,$^{5,14}$
M.~Symeonidis,$^{11,15}$
M.~Viero,$^{5}$
C.K.~Xu$^{5,14}$ and
M.~Zemcov$^{5,6}$}\vspace{0.4cm}\\
\parbox{\textwidth}{\raggedright
$^{\dagger}$ Herschel is an ESA space
    observatory with science instruments provided by European-led
    Principal Investigator consortia and with important participation
    from NASA.\\
$^{1}$Aix-Marseille Universit\'e, CNRS, LAM (Laboratoire d'Astrophysique de Marseille) UMR7326, 13388, France\\
$^{2}$Department of Astronomy, University of Maryland, College Park, MD 20742-2421, USA\\
$^{3}$Laboratoire AIM-Paris-Saclay, CEA/DSM/Irfu - CNRS - Universit\'e Paris Diderot, CE-Saclay, pt courrier 131, F-91191 Gif-sur-Yvette, France\\
$^{4}$Institut d'Astrophysique Spatiale (IAS), b\^atiment 121, Universit\'e Paris-Sud 11 and CNRS (UMR 8617), 91405 Orsay, France\\
$^{5}$California Institute of Technology, 1200 E. California Blvd., Pasadena, CA 91125, USA\\
$^{6}$Jet Propulsion Laboratory, 4800 Oak Grove Drive, Pasadena, CA 91109, USA\\
$^{7}$Center for Astrophysics and Space Astronomy 389-UCB, University of Colorado, Boulder, CO 80309, USA\\
$^{8}$Dept. of Physics \& Astronomy, University of California, Irvine, CA 92697, USA\\
$^{9}$Department of Physics, Virginia Tech, Blacksburg, VA 24061, USA\\
$^{10}$Department of Physics \& Astronomy, University of British Columbia, 6224 Agricultural Road, Vancouver, BC V6T~1Z1, Canada\\
$^{11}$Astronomy Centre, Dept. of Physics \& Astronomy, University of Sussex, Brighton BN1 9QH, UK\\
$^{12}$RAL Space, Rutherford Appleton Laboratory, Chilton, Didcot, Oxfordshire OX11 0QX, UK\\
$^{13}$Department of Astrophysics, Denys Wilkinson Building, University of Oxford, Keble Road, Oxford OX1 3RH, UK\\
$^{14}$Infrared Processing and Analysis Center, MS 100-22, California Institute of Technology, JPL, Pasadena, CA 91125, USA\\
$^{15}$Mullard Space Science Laboratory, University College London, Holmbury St. Mary, Dorking, Surrey RH5 6NT, UK}}
\begin{document}

\date{}

\maketitle

\label{firstpage}

\begin{abstract}
  We study the link between observed ultraviolet luminosity, stellar
  mass, and dust attenuation within rest-frame UV-selected samples at
  $z\sim 4, 3$, and 1.5. We measure by stacking at 250, 350, and
  500\,$\mu$m in the Herschel/SPIRE images from the HerMES program the
  average infrared luminosity as a function of stellar mass and UV
  luminosity. We find that dust attenuation is mostly correlated with
  stellar mass. There is also a secondary dependence with UV
  luminosity: at a given UV luminosity, dust attenuation increases
  with stellar mass, while at a given stellar mass it decreases with
  UV luminosity. We provide new empirical recipes to correct for dust
  attenuation given the observed UV luminosity and the stellar
  mass. Our results also enable us to put new constraints on the
  average relation between star formation rate and stellar mass at
  $z\sim 4, 3$, and 1.5. The star formation rate-stellar mass relations
  are well described by power laws ($\rmn{SFR}\propto M_*^{0.7}$),
  with the amplitudes being similar at $z\sim4$ and $z\sim3$, and
  decreasing by a factor of 4 at $z\sim1.5$ at a given stellar
  mass. We further investigate the evolution with redshift of the
  specific star formation rate. Our results are in the upper range of
  previous measurements, in particular at $z\sim3$, and are consistent
  with a plateau at $3<z<4$. Current model predictions
  (either analytic, semi-analytic or hydrodynamic) are inconsistent
  with these values, as they yield lower predictions than the
  observations in the redshift range we explore. We use these results
  to discuss the star formation histories of galaxies in the framework
  of the Main Sequence of star-forming galaxies. Our results suggest
  that galaxies at high redshift ($2.5<z<4$) stay around 1\,Gyr on the
  Main Sequence. With decreasing redshift, this time increases such
  that $z=1$ Main Sequence galaxies with $10^{8}<M_*/\msun<10^{10}$
  stay on the Main Sequence until $z=0$.
\end{abstract}



\section{Introduction}

Star formation is one the most important processes in galaxies, yet
our understanding of it is far from satisfactory. While it is commonly
recognised that the evolution of the large-scale structure of the
Universe is linked to that of dark matter, which is driven by
gravitation, baryonic physics is much more challenging. Having a good
understanding of star formation would be a great piece to put in the
puzzle of galaxy formation and evolution. The first step is to be able
to measure accurately the amount of star formation itself for a large
number of galaxies. This means we need to be able to build statistical
samples with observables that are linked to the recent star formation
activity. One of the easiest way to perform this is to consider
rest-frame ultraviolet (UV) selected samples, as the emission of
galaxies in this range of the spectrum is dominated by young,
short-lived, massive stars \citep{Kennicutt_1998}. Thanks to the
combination of various observatories, building UV-selected samples is
now feasible over most of the evolution of the Universe, from
$z\sim10$ to $z=0$ \citep[e.g.][]{Bouwens_2012, Ellis_2013,
  Martin_2005, Reddy_2012b}. There is however one drawback to this
approach, which is that the attenuation by dust is particularly
efficient in the UV \citep[e.g.][]{Calzetti_1997}. As the absorbed
energy is re-emitted in the far-infrared (FIR) range of the spectrum,
it is necessary to combine both of these tracers to get the complete
energy budget of star formation. The current observational facilities
however are such that it is much easier to build large samples from
the restframe UV than from the restframe IR over a wide redshift
range. It is then useful to look at the FIR properties of UV-selected
galaxies as a function of redshift in order to understand the biases
inherent to a UV selection, to characterise for instance the galaxy
populations probed by IR and UV selections, determine the amount of
total cosmic star formation rate probed by a rest-frame UV selection,
or the link between the level of dust attenuation \citep[as probed by
the ratio of IR to UV luminosities,][]{Gordon_2000} and physical
properties. This approach has been successful by combining UV
selections and \textit{Spitzer} data at $z\la 1$ to study the link
between dust attenuation and UV luminosity or stellar mass
\citep{Buat_2009, Martin_2007, Xu_2007}, as well as correlation with galaxy colors \citep{Arnouts_2013}. By measuring the ratio
between the cosmic star formation rate density estimated from IR and
UV selections, \citep{Takeuchi_2005} showed that the fraction of the
cosmic star formation rate probed by a UV selection, without
correction for dust attenuation, decreases from 50 per cent to 16 per
cent between $z=0$ and $z=1$ \citet{Takeuchi_2005}. At $z>1.5$,
\textit{Spitzer} data probe the mid-IR range of the spectrum, which
can lead to an overestimation of the IR luminosity
\citep[e.g.][]{Elbaz_2010}. At these redshifts, \textit{Herschel}
\citep{Pilbratt_2010} data become particularly valuable for such
projects. \citet{Reddy_2012a} extended this kind of study by stacking
$z=2$ Lyman Break Galaxies (LBGs) in \textit{Herschel}/PACS
\citep{Poglitsch_2010} images to investigate their dust attenuation
properties: they estimated that typical UV-selected galaxies at these
epochs have infrared luminosities similar to Luminous Infrared
Galaxies (LIRGs, $10^{11}<L_{\rm
  IR}/\lsun<10^{12}$). \citet{Burgarella_2013} combined the
measurements at $0<z<4$ of the UV \citep{Cucciati_2012} and IR
\citep{Gruppioni_2013} restframe luminosity functions to infer the
redshift evolution of the total (UV+IR) cosmic star formation rate and
dust attenuation. In a previous study based on a stacking analysis of
UV-selected galaxies at $z\sim1.5$ in \textit{Herschel}/SPIRE
\citep{Griffin_2010} images, we showed that using a UV-selection at
$z\sim1.5$ with a proper correction for dust attenuation enables us to
recover most of the total cosmic star formation activity at that epoch
\citep{Heinis_2013}.

It is also necessary to investigate the link between dust attenuation
and a number of galaxy properties, in order to be able to accurately
correct for dust attenuation, by providing empirical relations for
instance. One of the most commonly used empirical relation in this
context is based on the correlation between the slope of the UV
continuum and the dust attenuation \citep{Meurer_1999}. Such
correlation has been observed for star-forming galaxies from high to
low redshifts \citep[e.g.][]{Buat_2005, Burgarella_2005, Heinis_2013,
  Reddy_2010, Seibert_2005}. However, the common assumption that the
relation derived from local starbursts \citep{Calzetti_2001,
  Meurer_1999} is universal is questionable \citep{Heinis_2013,
  Hao_2011} as the extinction curve is dependent on the dust geometry
\citep[e.g.][]{Calzetti_2001} and dust properties
\citep[e.g.][]{Inoue_2006}. Moreover, the UV slope of the continuum
encodes partly the star formation history of the galaxies
\citep{Boquien_2012, Kong_2004, Panuzzo_2007}, and the observed
relation between the UV slope and the dust attenuation is also
selection-dependent \citep{Buat_2005, Seibert_2005}.

It is then useful to turn towards other observables, which might
provide better ways to correct for dust attenuation in a statistical
sense. Dust attenuation is for instance not really well correlated
with observed UV luminosity \citep[e.g.][]{Buat_2009,Heinis_2013,
  Xu_2007}. On the other hand, the correlation with stellar mass is
tighter \citep[e.g.][]{Buat_2012, Finkelstein_2012, Garn_2010,
  Pannella_2009, Xu_2007}. This is somewhat expected as the dust
production is linked to the star formation history, through heavy
elements production, and stellar mass in this context can be seen as a
crude summary of star formation history.

Investigating the link between dust attenuation and stellar mass is
interesting by itself, but getting a direct estimate of the IR
luminosity implies that we can also derive the star formation rate
(SFR) accurately. This means that we are able for instance to
characterise the relation between the SFR and the stellar mass. By
considering galaxy samples based on star-formation activity, we are
actually expecting to deal with objects belonging to the so-called
`Main Sequence' of galaxies. A number of studies pointed out that
there is a tight relation between the SFR and the stellar mass of
galaxies, from high to low redshift \citep[][]{Bouwens_2012,
  Daddi_2007, Elbaz_2007, Noeske_2007, Wuyts_2011b}. Galaxies on this
Main Sequence are more extended than starbursts \citep{Elbaz_2011,
  Farrah_2008, Rujopakarn_2013}, the latter representing only a small
contribution, in terms of number density, to the global population of
star forming galaxies \citep{Rodighiero_2011}. The relation between
SFR and stellar mass also seems to be independent of the environment
of the galaxies \citep{Koyama_2013}.  While there is debate on the
slope and scatter of this relation, it is definitely observed at
various redshifts, with its amplitude decreasing with cosmic time
\citep{Iglesias-Paramo_2007, Martin_2007, Noeske_2007,
  Wuyts_2011b}. The mere existence of this relation raises a number of
issues for galaxy formation and evolution, as it implies that galaxies
experience a rather smooth star formation history.

In this paper, we take advantage of the combination of the
multiwavelength data available within the COSMOS field
\citep{Scoville_2007}, with the \textit{Herschel}/SPIRE observations
obtained in the framework of the Herschel Multi-Tiered Extragalactic
Survey key program\footnote{\url{http://hermes.sussex.ac.uk}}
\citep[HerMES,][]{Oliver_2012}. We are assuming here that the
rest-frame FIR emission we measure originates from the dust
responsible for the UV/optical attenuation. Indeed, the wavelength
range covered by SPIRE is dominated by the emission of dust heated by
stars, the contribution from dust heated by Active Galactic Nuclei
being significantly lower at these wavelengths
\citep{Hatziminaoglou_2010}. Moreover, our UV-selection biases against
galaxies dominated by old stellar populations, hence the FIR emission
we measure is mostly due to the dust heated by young stellar
populations.

We focus on three UV-selected samples at $z\sim4,3$, and 1.5 \citep[see][ for a similar study based on H$\alpha$-selected sample at $z=1.47$]{Ibar_2013}. We
revisit the relations between dust attenuation and UV luminosity as
well as stellar mass, over this wide redshift range, using homogeneous
selections and stellar mass determination. Our aim is to disentangle
the link between dust attenuation and these two physical quantities,
by directly measuring their IR luminosities thanks to
\textit{Herschel}/SPIRE data. We also put new constraints on the
SFR-stellar mass relations from $z\sim4$ to $1.5$, and use our results
to discuss the star formation histories of Main Sequence galaxies.

This paper is organised as follows: in Sect. \ref{sec_data} we present
the UV-selected samples we build from the multiwavelength data
available in the COSMOS field. As most of the galaxies of these
samples are not detected individually with \textit{Herschel}/SPIRE, we
perform a stacking analysis, and describe the methods we use in
Sect. \ref{sec_stacking}. We present our results in
Sect. \ref{sec_results}: we detail the relations between dust
attenuation and UV luminosity (Sect. \ref{sec_irx_luv}) and between
dust attenuation and stellar mass (Sect. \ref{sec_irx_mass}). We
present in Sect. \ref{sec_sfr_mass} the SFR-stellar mass relations for
UV-selected samples we obtain at $z\sim1.5, 3$ and 4. We also investigate
the link between dust attenuation and UV luminosity and stellar mass
jointly (Sect. \ref{sec_irx_luv_mass}). We discuss these results in
Sect. \ref{sec_discussion} and present our conclusions in
Sect. \ref{sec_conclusion}.



Throughout this paper, we make the following assumptions: we use a
standard cosmoslogy with $\Omega_{\rm m} = 0.3$, $\Omega_{\Lambda} =
0.7$, and $H_0 = 70\,$km s$^{-1}$ Mpc$^{-1}$; we denote far-UV (FUV)
and IR luminosities as $\nu L_{\nu}$; use AB magnitudes, and consider
a \citet{Chabrier_2003} Initial Mass Function (IMF). When comparing to
other studies, we consider that no conversion is needed for SFR and
stellar mass estimates between \citet{ Kroupa_2001} and
\citet{Chabrier_2003} IMFs. When converting from \citet{Salpeter_1955}
IMF to \citet{Chabrier_2003} IMF, we divide $M_{*, \rm Salpeter}$ by
1.74 \citep{Ilbert_2010}, and SFR$_{\rm Salpeter}$ by 1.58
\citep{Salim_2007}.



\section[]{Data samples}\label{sec_data}
\begin{table}
\begin{minipage}{100mm}
\caption{Description of UV-selected samples}
\label{tab_samples}
\begin{tabular}{@{}lccc}
\hline
 & \multicolumn{3}{c}{Sample}\\
 & $z\sim1.5$ & $z\sim3$ & $z\sim4$\\
\hline
mag. limit\footnote{Magnitude limit of the sample} &$u^*=26$ &$r^+=26$ & $i^+=26$\\[0.1cm]
$z_{\rm phot\, range}$\footnote{Used range of photometric redshifts} &$1.2-1.7$ &$2.75-3.25$ &$3.5-4$ \\[0.1cm]
$\langle z_{\rm phot} \rangle$\footnote{Mean photometric redshift} &1.43 &2.96 &3.7 \\[0.1cm]
$\left\langle \sigma(z_{\rm phot}) \right\rangle$\footnote{Mean photometric redshift error, in $1+z$} & 0.04&0.1 &0.17 \\[0.1cm]
$N_{\rm gal}$ &42,184 &23,774 & 7,713\\[0.1cm]
$\lambda_{\rm rest\, eff}$[\AA]\footnote{Effective restframe wavelength \citep[from][]{Ilbert_2009} at mean redshift}&1609 &1574 & 1623\\[0.1cm]
$\log(L_{\rm FUV}(\langle z_{\rm phot} \rangle,\rm{mag. limit}) [L_{\sun}])$\footnote{FUV luminosity at mean redshift and magnitude limit of the sample} & 9.6 & 10.1 &10.3 \\[0.1cm]
$\left\langle \sigma\left(\log(M_* [\msun])\right) \right\rangle$\footnote{Mean stellar mass error} &0.15 &0.27 &0.30 \\[0.1cm]
$\log(M_* [\msun])$ reliability limit\footnote{Reliability limit in stellar mass (see Sect. \ref{sec_uv_sel})} & 9.5 &10.3 & 10.6\\[0.1cm]
\hline
\end{tabular}
\end{minipage}
\end{table}

\subsection{Photometric redshifts and stellar masses}
We base this study on the photometric redshift catalogue built from the
COSMOS data by \citet[][version 2.0]{Ilbert_2009}. This catalogue is based on an $i$-
band detection, down to 0.6$\,\sigma$ above the background \citep{Capak_2007}. These estimates
benefit from new near-infrared imaging in the $Y$, $J$, $H$, and $K_s$ bands obtained
with the VISTA telescope as part of the UltraVISTA project
\citep{McCracken_2012}. In the redshift range $1.5<z<3.5$, the
precision on the photometric redshifts (defined as the scatter of the difference with spectroscopic redshifts, in 1+$z$) is around 3 per cent. This value is given
by \citet{Ilbert_2013} for objects with $K_s<24$, and has been obtained by comparing to zCOSMOS
faint sample ($I_{med} =23.6$) and faint DEIMOS spectroscopic redshifts ($I_{mean}=23.5$). At $z
\sim4$, the spectroscopic redshifts available ($I_{med} =24.4$) yield a precision of 4 per cents, and
suggest that the contamination from low redshift galaxies is negligible. On the other hand, this
spectroscopic sample at $z\sim4$ is not likely to be representative of our sample at the same redshifts (see Table \ref{tab_samples}). 
The actual photometric redshift error for our samples might be larger than this, as we are dealing with
 fainter objects. We also quote in Table \ref{tab_samples} as an alternative the mean photometric
redshift error, in (1+z), estimated from the PDF of the photometric redshifts derived by
 \citet{Ilbert_2009}. \citet{Ilbert_2010} showed that the error measured from the PDF
is a robust estimate of the accuracy as measured with respect to
spectroscopic objects. At $z\sim3$, the mean error from the PDF is 0.1, and 0.17 at $z\sim4$.

We also consider in this paper the stellar masses estimates of
\citet[][version 2.0]{Ilbert_2009}. Briefly, the stellar masses are derived from SED
fitting to the available photometry, assuming \citet{Bruzual_2003}
single stellar population templates, an exponentially declining star
formation history, and the \citet{Chabrier_2003}
IMF. \citet{Ilbert_2013} showed that the assumption of an
exponentially declining star formation history does not have a strong
impact on the stellar masses estimates.

\subsection{UV-selected samples}\label{sec_uv_sel}
We consider three UV-selected samples at $z\sim1.5$, $z\sim 3$, and
$z\sim 4$. The sample at $z\sim1.5$ has already been presented in
\citet{Heinis_2013}. We detail here how we build the samples at $z\sim
3$, and $z\sim 4$. We use optical imaging of the COSMOS field from
\citet{Capak_2007} in $r^+$ and $i^+$, both from Subaru. We
cross-match single band catalogues built from these images with the
photometric redshift catalogue of \citet[][version 2.0]{Ilbert_2009}. Ninety-nine per cent of the
objects with $u^*<26$ have a counterpart in the catalogue of \citet[][version 2.0]{Ilbert_2009}, while 92
per cent of objects with $r^+<26$ have a counterpart. In the $i^+$-band, we use directly the catalog of
\citet[][version 2.0]{Ilbert_2009}, as it is based on an $i^+$-band detection.

We then build
UV-selected samples, at $z\sim 3$ and $z\sim 4$. We detail in Table
\ref{tab_samples} the main characteristics of the three samples we
consider here. All these samples probe the FUV rest-frame range of the
spectrum, with rest-frame effective wavelengths within the range
$1570-1620$\,\AA~at the mean redshifts of the samples (see Table
\ref{tab_samples}).

We will perform stacking at 250, 350 and 500\,$\mu$m as a function of
FUV luminosity, $L_{\rm FUV}$, and stellar mass $M_*$. We derive
$L_{\rm FUV}$ from the observed magnitude as follows:
\begin{equation}\label{eq_m2luv}
  L_{\nu} = \frac{4\pi D_{L}^2(z)10^{-0.4(48.6+m)}}{1+z}
\end{equation}

where $D_{L}(z)$ is the luminosity distance at $z$, and $m$ is the
observed magnitude: we use $u^*$ at $z\sim1.5$, $r^+$ at $z\sim 3$,
and $i^+$ at $z\sim 4$. We then compute the UV luminosity at
1530\,\AA.

We estimate a reliability limit in stellar mass for each sample the
following way. We compute, as a function of $M_*$, the fraction of
objects with 3.6\,$\mu$m flux measurements fainter than the 80 per
cent completeness limit \citep[2.5\,$\mu$Jy,][]{Ilbert_2010}. We
choose the reliability limit as the minimum $M_*$ value where this
fraction is lower than 0.3. In other words, above this value of $M_*$,
the fraction of objects that have a flux at 3.6\,$\mu$m larger than
the 80\% completeness limit is $\ge0.7$. Note that we do not impose a
cut on 3.6\,$\mu$m fluxes. The stellar mass is also estimated for
objects with 3.6\,$\mu$m flux fainter than 2.5\,$\mu$Jy, however this
estimate is less robust than for brighter objects. We quote the reliability limits for each sample in Table \ref{tab_samples}.




\section{Stacking measurements}\label{sec_stacking}
We base our study on the \textit{Herschel}/SPIRE imaging of the COSMOS
field obtained within the framework of the HerMES key
program\citep{Oliver_2012}. Most of the objects from our UV-selected
samples are not detected individually in these images, so we rely on a
stacking analysis. We use the same methods as those presented in
\citet{Heinis_2013} to measure flux densities using
stacking\footnote{We stack here in flux rather than in luminosity
  \citep[e.g.][]{Oliver_2010, Page_2012}. The latter requires to
  estimate beforehand the kcorrection in the IR, which would be not
  reliable for most of our objects, not detected at shorter IR
  wavelengths.}. We recall here only the main characteristics of the
methods. We perform stacking using the IAS library
\citep{Bavouzet_2008,
  Bethermin_2010}\footnote{\url{http://www.ias.u-psud.fr/irgalaxies/files/ias_stacking_lib.tgz}}. We
use mean stacking, without cleaning images from detected sources. We
showed in \citet{Heinis_2013} that using our method or median stacking
with cleaning images from detected sources, yields similar results. We
correct the stacking measurements for stacking bias, using extensive
simulations of the detection process of the sources. We perform these
simulations by injecting resolved artificial sources in the original
images, and keeping track of the recovered sources. We then use the
stacking of these artificial sources to correct the actual
measurements. We also correct for the clustering of the input
catalogue by taking into account the angular correlation function of
the input sample.

We derive errors on the stacking flux densities by bootstrap
resampling. We use hereafter the ratio of the stacking flux density
over its error as a measurement of signal-to-noise ratio. For each
stacking measurement, we obtain a flux density at 250, 350 and
500\,$\mu$m. We derive an infrared luminosity $L_{\rm IR}$ by
adjusting these fluxes to the \citet{Dale_2002} templates, using the
SED-fitting code CIGALE\footnote{\url{http://cigale.oamp.fr/}}
\citep{Noll_2009}. The \citet{Dale_2002} templates have been shown to
be a reasonable approximation of the SEDs of \textit{Herschel} sources
\citep{Elbaz_2010, Elbaz_2011}. We consider \lir as the integration of
the SED over the range $8<\lambda<1000\,\mu$m. CIGALE estimates the
probability distribution function of \lirns. We consider the mean of
this distribution as our \lir value, and the standard deviation as the
error on \lirns. We use as redshift the mean redshift of the galaxies
in the bin.

Hereafter, we perform stacking as a function of \luv and $M_*$
separately in Sect. \ref{sec_lir_luv}, \ref{sec_irx_luv},
\ref{sec_irx_mass}, and \ref{sec_sfr_mass}, and we also perform
stacking as a function of both \luv and $M_*$ in
Sect. \ref{sec_irx_luv_mass}. We characterise each bin by the mean
value of \luv and/or $M_*$. We derive the errors on the mean \luv
using mock catalogues. These mock catalogues are only used to estimate
errors on mean \luv and $M_*$. We build 100 mock catalogues, with new
redshifts for each object, drawn within the probability distribution
functions derived by \citet{Ilbert_2010}. We can then assign new \luv
using eq. \ref{eq_m2luv}. For a given stacking measurement including a
given set of objects, we compute the mean of \luv for each mock
catalogue. The error on the mean $L_{\rm FUV}$ is then the standard
deviation of the means obtained from all mock catalogues. We derive
errors on the mean $M_*$ in a similar way, using the stellar mass
probability distribution functions derived by \citet{Ilbert_2010}.

\section{Results}\label{sec_results}

We first show results of the stacking as a function of \luvns; we look
at the relation between the average \lir and \luv
(Sect. \ref{sec_lir_luv}) and then at the relation between the dust
attenuation, probed by the IR to UV luminosity ratio, and \luv
(Sect. \ref{sec_irx_luv}). 

We further turn to results we obtain by stacking as a function of
stellar mass, looking at the relation between dust attenuation and
stellar mass (Sect. \ref{sec_irx_mass}). We also investigate the joint
dependence between \luvns, $M_*$, and dust attenuation
(Sect. \ref{sec_irx_luv_mass}).

As we obtain estimates of $L_{\rmn{IR}}$, we derive a total star
formation rate by combining with the observed UV luminosity, and look
at the relation between star formation rate and stellar mass in our
samples (Sect. \ref{sec_sfr_mass}).

\subsection{Stacking as a function of \luv}

\subsubsection{\lir-\luv relation from $z\sim4$ to $z\sim1.5$}\label{sec_lir_luv}
\begin{figure}
\includegraphics[width=\hsize]{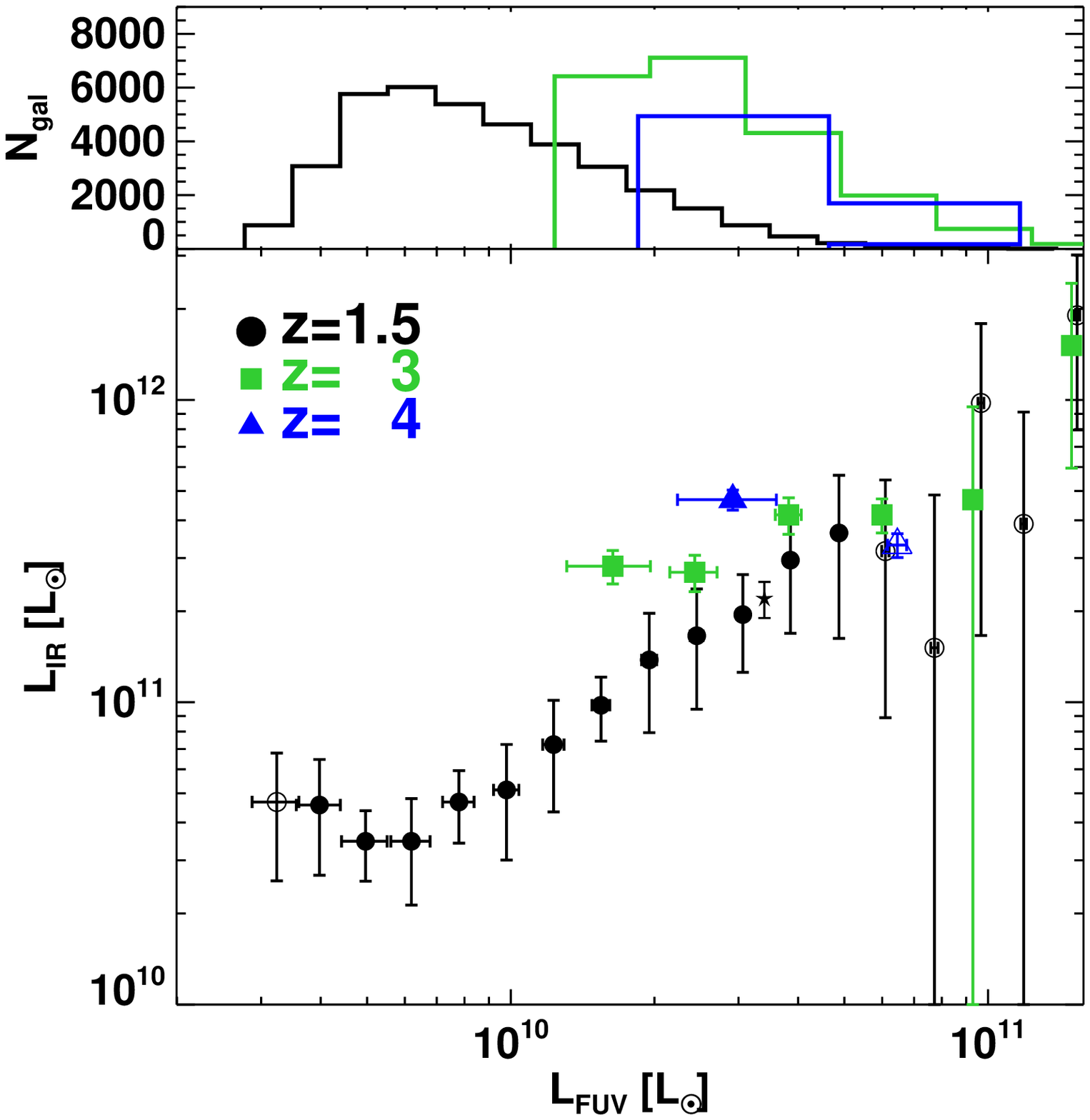}
\caption{\textit{Top:} histogram of the number of galaxies included in
  the stacking measurements at $z\sim1.5$ (black), $z\sim3$ (green),
  and $z\sim4$ (blue). \textit{Bottom:} \lir versus \luvns, at
  $z\sim1.5$ \citep[from][black circles]{Heinis_2013}, $z\sim3$ (green
  squares), and $z\sim 4$ (blue triangles). Filled symbols represent
  stacking measurements with $S/N>3$ in the three SPIRE bands, and
  open symbols measurements which do not meet this criterion. The star
  symbol show the result of stacking measurements by \citet{Reddy_2012a}
  on a sample of Lyman Break Galaxies at $z=2$.}
\label{fig_stack_lfuv_vs_lir}
\end{figure}

In Fig. \ref{fig_stack_lfuv_vs_lir}, we show the \lir measured by
stacking as a function of \luv at $z\sim1.5, 3,$ and 4. At $z\sim1.5$,
for galaxies with $3\times10^{9}<L_{\rm FUV}/\lsun<8\times10^{9}$, \lir is
roughly constant at $L_{\rm IR}\sim 4\times10^{10}\lsun$. For \luv
brighter than $8\times10^{9}\,\lsun$, \lir is increasing with \luvns, with a
power law slope of $1.1\pm0.2$. This shows that in this range of UV
luminosities at $z\sim1.5$, \lir and \luv are well correlated.

At $z\sim3$ and $z\sim4$, the situation is quite different. At these
redshifts, we explore a smaller dynamic range of UV luminosities,
$10^{10}<L_{\rm FUV}/\lsun<10^{11}$. At these epochs, we do not
measure any statistically significant trend of \lir with \luv in
UV-selected samples. We find that \lir is roughly constant at $L_{\rm
  IR} \sim 4\times 10^{11}\lsun$.






\subsubsection{Dust attenuation as a function of \luv from $z\sim4$ to
  $z\sim1.5$}\label{sec_irx_luv}

\begin{figure}
\includegraphics[width=\hsize]{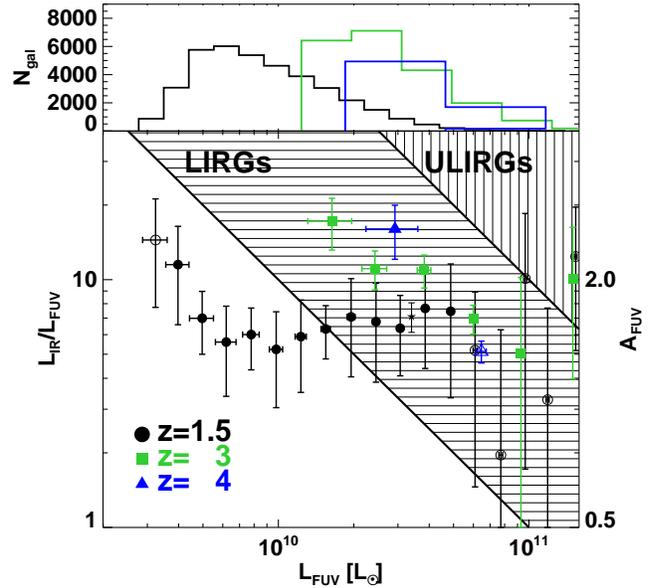}
\caption{\textit{Top:} histogram of the number of galaxies included in
  the stacking measurements at $z\sim1.5$ (black), $z\sim3$ (green),
  and $z\sim4$ (blue). \textit{Bottom:} IR to UV luminosity ratio
  versus \luvns, at $z\sim1.5$ \citep[from][black
    circles]{Heinis_2013}, $z\sim3$ (green squares), and $z\sim 4$
  (blue triangles). Filled symbols represent stacking measurements
  with $S/N>3$ in the three SPIRE bands, and open symbols measurements
  which do not meet this criterion. The right axis shows the
  equivalent attenuation in the FUV band, in magnitudes, using
  eq. \ref{eq_afuv}. The horizontally hatched area represents the
  region where LIRGs are, and the vertically hatched area represents
  the same for ULIRGs. The star symbol show the stacking results of
  \citet{Reddy_2012a} on a sample of Lyman Break Galaxies at $z=2$.}
\label{fig_stack_lfuv}
\end{figure}

In Fig. \ref{fig_stack_lfuv}, we show the relations between the
$L_{\rm IR}/L_{\rm FUV}$ ratio, a proxy for dust attenuation, and \luv
at $z\sim3$ and $z\sim4$. We also show for comparison the results we
obtained at $z\sim1.5$ \citep{Heinis_2013}.

We indicate the equivalent dust attenuation in the FUV, $A_{\rm FUV}$,
derived from the \lir to \luv ratio using \citep{Buat_2005}:

\begin{eqnarray}\label{eq_afuv}
  A_{\rm FUV} & = &-0.0333\rm{IRX}^3 + 0.3522\rm{IRX}^2+1.1960\rm{IRX} \nonumber\\
            & + & 0.4967\\
  \rm{IRX}       & = & \log\left(\frac{L_{\rm IR}}{L_{\rm FUV}}\right). \nonumber
\end{eqnarray}

In the ranges of UV luminosity we probe, the relations between dust
attenuation and \luv change from $z\sim4$ to $z\sim1.5$. At $z\sim1.5$, the
dust attenuation is mostly independent of \luvns. At $z\sim3$ and $z\sim4$,
we observe that the dust attenuation on average decreases with
\luvns. This decrease is linked to the fact that \luv is not well
correlated with \lirns, as suggested by
Fig. \ref{fig_stack_lfuv_vs_lir}.



Our results also show that at given \luvns, dust attenuation is larger
at $z\sim3,4$ than at $z\sim1.5$ for galaxies with $L_{\rm FUV}
<4\times 10^{10} L_{\odot}$. We show later that this effect is
actually linked to the stellar mass of the galaxies (see
Sect. \ref{sec_irx_luv_mass}).

\subsection[]{Dust attenuation as a function of stellar mass}\label{sec_irx_mass}
\begin{figure}
\includegraphics[width=\hsize]{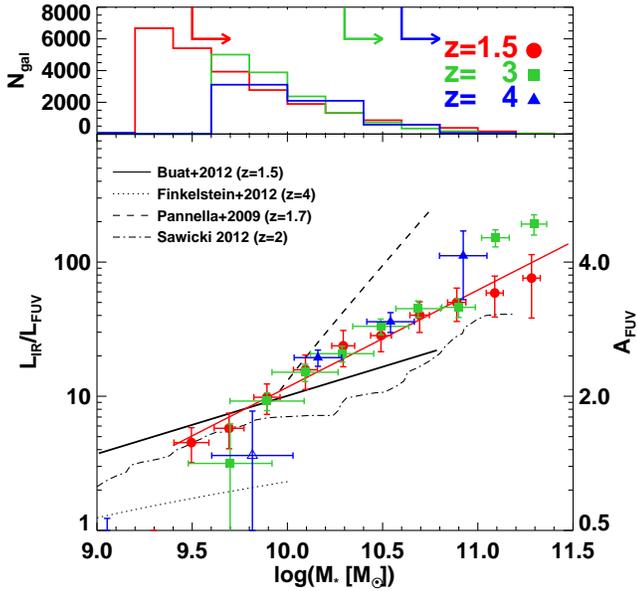}
\caption{\textit{Top:} histogram of the number of galaxies included in
  the stacking measurements at $z\sim1.5$ (red), $z\sim3$ (green), and
  $z\sim4$ (blue). The arrows show the mass reliability limits for
  each sample. \textit{Bottom:} IR to UV luminosity ratio versus
  stellar mass, at $z\sim1.5$ (red circles), $z\sim3$ (green squares),
  and $z\sim 4$ (blue triangles). The right axis shows the equivalent
  attenuation in the FUV band, in magnitudes, using
  eq. \ref{eq_afuv}. The solid red line shows our fit to the
  $z\sim1.5$ measurements. The various black lines show previous
  measurements at various redshifts from \citet[][$z=1.5$]{Buat_2012},
  \citet[][$z=4$]{Finkelstein_2012} (whose results agree really well
  with \citet[][$z=4$]{Bouwens_2012}, that we do not show here),
  \citet[][$z=2$]{Pannella_2009}, and \citet[][$z=2$]{Sawicki_2012}.}
\label{fig_stack_M*}
\end{figure}

We investigate here the relation between dust attenuation and stellar
mass. We show in Fig. \ref{fig_stack_M*} our measurements of the ratio of
IR to UV luminosity as a function of stellar mass, at $z\sim4$,
$z\sim3$, and $z\sim1.5$.

The link between dust attenuation and stellar mass is strikingly
different from the link between dust attenuation and UV luminosity. At
all the redshifts we consider here, there is a clear correlation, on
average, between dust attenuation and stellar mass. The results in
Fig. \ref{fig_stack_M*} show that the $L_{IR}/L_{FUV}$ ratio is much
better correlated with stellar mass than with UV luminosity. Within
the same samples, the $L_{IR}/L_{FUV}$ ratio varies by a factor of two
at most as a function of \luvns, while it varies by one order of
magnitude as a function of $M_*$. Our results also suggest that there
is no significant evolution with redshift of the dust attenuation at a
given stellar mass, between $z\sim4$ and $z\sim1.5$. There is a
possible trend at the high mass range ($M_*>10^{11}\msun$) that dust
attenuation decreases between $z\sim3$ and $z\sim1.5$. The statistics
is however low for these mass bins, and the fraction of UV-selected
objects directly detected at SPIRE wavelengths is the highest.

Assuming that the relation between the \lir to \luv ratio and $M_*$
can be parameterised as:

\begin{equation}
  \textrm{IRX} = \alpha\log\left(\frac{M_*}{10^{10.35}}\right) + \textrm{IRX}_0
\end{equation}

we obtain as best fits parameters at $z\sim1.5$ $\alpha=0.72\pm0.08$
and $\textrm{IRX}_0 = 1.32\pm0.04$.  This relation is valid at
$z\sim1.5, 3, $ and 4 for $10^{10}<M_*/\msun<10^{11}$.

We compare our results with previous estimates of the relation between
dust attenuation and stellar mass for UV-selected samples. At $z\sim
1.5$, our results are in reasonable agreement with those from
\citet{Buat_2012}, derived from SED fitting, based on UV-selected
objects with spectroscopic redshifts and photometry from the restframe
UV to the restframe FIR. Our results are also in good agreement with those from
\citet{Whitaker_2012} at $1.<z<1.5$, who studied a mass-selected sample of star-forming galaxies.
 Our findings are also consistent with those from \citet{Wuyts_2011b}, who observed that the ratio of
 SFRs derived from the IR and the UV increases with total SFR ($=$SFR$_{\rmn{IR}} + $SFR
 $_{\rmn{UV}}$) and $M_*$. While we observe a higher amplitude at a
given mass, our measurements show a slope of the IRX$-M_*$ relation
similar to the one derived by \citet{Sawicki_2012}, whose results are
derived from SED fitting applied to a sample of BX galaxies at $z\sim
2.3$, using photometric redshifts, and UV/optical restframe data. We
also compare our results at $z\sim4$ with the measurements of
\citet{Finkelstein_2012}, who studied the link between the slope of
the UV continuum, $\beta$, and the stellar mass. We converted their
measurements of $\beta$ to $A_{\rm FUV}$ assuming the
\citet{Meurer_1999} relation, which has been claimed to be valid at
$z=4$ \citep{Lee_2012}. The measurements of \citet{Finkelstein_2012}
probe a lower mass range than ours, making a direct comparison
difficult. Our measurement in the lowest mass bin we probe at $z\sim4$
is in formal agreement with theirs, however it has a low signal to
noise ratio, and may suffer from significant incompleteness in mass as
well. Nevertheless, the extrapolation of the relation observed by
\citet{Finkelstein_2012} at higher masses does not match our
measurements. We also compare our results with the relation derived by
\citet{Pannella_2009} at $z=2$, from radio stacking of a sample of
BzK-selected galaxies. This relation would significantly overpredict
the dust attenuation for a UV-selected sample when compared to our
results. These different relations between dust attenuation and
stellar mass for UV and BzK-selected samples coud be due to the fact
that the BzK selection is less sensitive to dust attenuation, and
probes galaxies that are dusty enough to be missed by UV selections
\citep[e.g.][]{Riguccini_2011}. We note the more recent results from
\citet{Pannella_2013} are in better agreement with our measurements.


\subsection{Dust attenuation as a function of stellar mass and UV luminosity}\label{sec_irx_luv_mass}
\begin{figure*}
\includegraphics[width=\hsize]{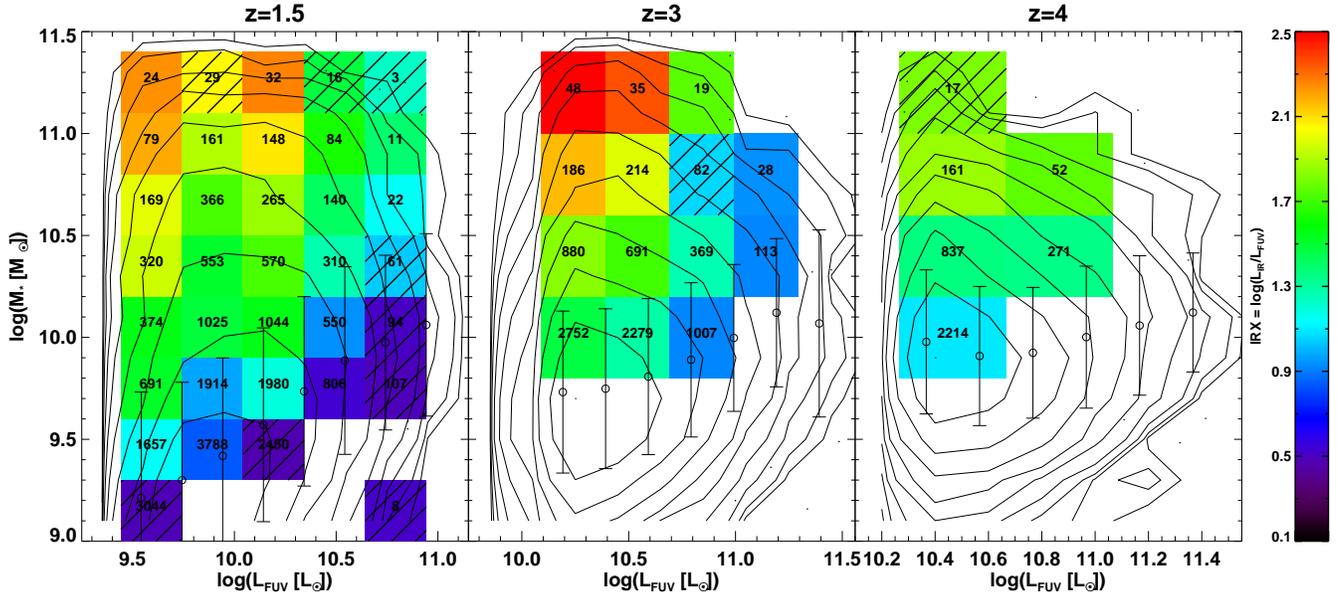}
\caption{Dust attenuation as a function of stellar mass and UV
  luminosity at $z\sim 1.5$ (left), $z\sim 3$ (middle) and $z\sim 4$
  (right). The color codes $\textrm{IRX} = \log(L_{\rm IR}/L_{\rm
    UV})$ in each cell where the stacking measurement is
  meaningful. Filled cells indicate stacking measurements with $S/N>3$
  in the 3 SPIRE bands, while black hatched cells show measurements
  with at most two SPIRE bands with $S/N>3$. The number of galaxies
  contributing to the stacking is indicated in each cell. The contours
  show the distribution of galaxies in the $(L_{\rm FUV}, M_*)$
  plane. The empty circles show the mean stellar mass for a given UV
  luminosity bin, with the dispersion as error bar.}
\label{fig_stack_mstar_luv}
\end{figure*}

The results presented in Sects. \ref{sec_irx_luv} and
\ref{sec_irx_mass} show that dust attenuation is on average well
correlated with stellar mass, and that this correlation is tighter
than the correlation between dust attenuation and \luvns. However,
dust attenuation is not \textit{completely} independent of \luv: while
at $z\sim1.5$, dust attenuation is mostly constant for $ 5 \times
10^{9}<L_{\rm FUV}/\lsun<5 \times 10^{10}$ it increases for fainter UV
luminosities. On top of this, dust attenuation is higher at $z\sim3$
than at $z\sim1.5$ at the same \luvns, but is found to be decreasing
with \luvns. It seems then that dust attenuation depends \textit{both}
on \luv and $M_*$, and that we need to investigate what is the link
between dust attenuation and these two quantities.

We performed stacking as a function of \luv and $M_*$ at $z\sim1.5,
3,$ and 4, using binnings of $\left(\Delta \log(L_{\rm FUV}/\lsun),
  \Delta \log(M_*/\msun)\right) = (0.3,0.3), (0.3,0.4)$, and
$(0.4,0.4)$ respectively. We show in Fig. \ref{fig_stack_mstar_luv}
the result of the stacking as a function of UV luminosity and stellar
mass. Note that filled cells indicate bins where the stacking
measurements have $S/N>3$ in all SPIRE bands, hatched cells bins where
there is at most two SPIRE band with $S/N>3$, and other cells are kept
empty. These empty cells indicate that there is no robust stacking
detection in these bins.

The measurements in Fig. \ref{fig_stack_mstar_luv} clearly show that
dust attenuation depends both on \luv and $M_*$. Dust attenuation
increases with $M_*$ at a given \luvns, while it decreases with \luv
at a given $M_*$. We already observed an increase of the dust
attenuation for faint UV galaxies \citep{Heinis_2013} at $z\sim1.5$
\citep[also observed previously by][]{Buat_2009, Buat_2012,
  Burgarella_2006}. Indeed, galaxies with large stellar masses and
strong dust attenuation exhibit faint UV luminosities, which is true
for all redshifts we study here. The results in
Fig. \ref{fig_stack_mstar_luv} also show that the range of dust
attenuation values over the stellar mass range decreases with \luvns,
as suggested in a previous study \citep{Heinis_2013}. We also
represent in Fig. \ref{fig_stack_mstar_luv} the location of the mean
stellar mass for each UV luminosity bin. The results at $z\sim 1.5$ in
particular show that lines of constant dust attenuation follow lines
roughly parallel to this relation. This explains the global lack of
dependence of dust attenuation with \luvns at $z\sim1.5$
\citep{Heinis_2013}. At $z\sim3$ and $z\sim4$, there is only a weak
correlation between $M_*$ and \luvns. This implies that bins in \luv
are mostly dominated by low mass galaxies in these samples. As shown
in Fig. \ref{fig_stack_mstar_luv}, the dust attenuation at a given
mass decreases with \luvns, which is exactly what we observe when
stacking as a function of \luv only.

The relation between dust attenuation and $(L_{\rm FUV},M_*)$ also
depends on redshift. Indeed, at a given mass and \luvns, the
attenuation is higher at $z\sim3$ than at $z\sim1.5$. For instance,
galaxies with $10^{10}\la L_{\rm FUV}/L_{\odot}\la 10^{10.35}$ have a
dust attenuation roughly 0.2 dex larger at a given mass at $z\sim3$
with respect to galaxies at $z\sim1.5$. This, combined with slightly
different \luv-$M_*$ relations explains why the dust attenuation for
this range of UV luminosities is larger at $z\sim3$ compared to
$z\sim1.5$ (see Fig. \ref{fig_stack_lfuv}).

We can use the results presented above in order to provide empirical
recipes to estimate dust attenuation as a function of $M_*$ and
\luvns. We detail those in Appendix \ref{app_emp_irx}.

\subsection{Star formation rate-stellar mass relations from $z\sim4$ to $z\sim1.5$}\label{sec_sfr_mass}


\begin{figure}
\includegraphics[width=\hsize]{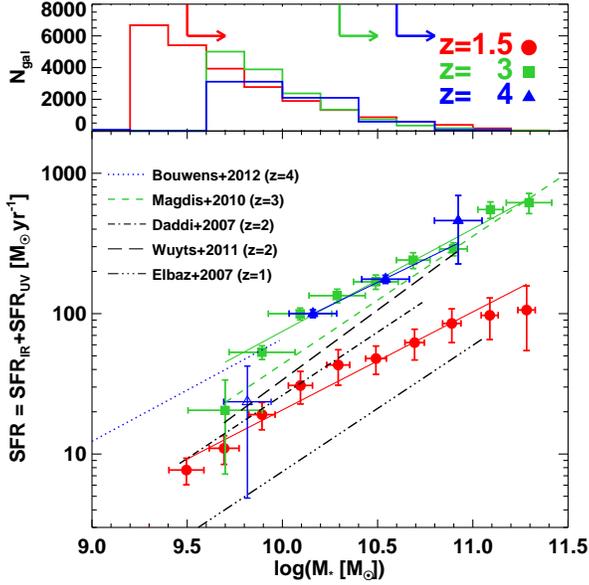}
\caption{\textit{Top:} histogram of the number of galaxies included in
  the stacking measurements at $z\sim1.5$ (red), $z\sim3$ (green), and
  $z\sim4$ (blue). The arrows show the mass reliability limits for
  each sample. \textit{Bottom:} star formation rate (sum of IR and FUV
  contributions) versus stellar mass, at $z\sim1.5$ (red circles),
  $z\sim3$ (green squares), and $z\sim 4$ (blue triangles). The
  various lines show previous measurements at various redshifts from
  \citet[][$z=4$]{Bouwens_2012}, \citet[][$z=2$]{Daddi_2007},
  \citet[][$z=1$]{Elbaz_2007}, \citet[][$z=2$]{Magdis_2010}, and
  \citet[][$z=2$]{Wuyts_2011b}.}
\label{fig_stack_sfr}
\end{figure}

\begin{table}
\caption{Fits to average star formation rate - stellar mass relations}
\label{tab_sfr_mass_fits}
\begin{tabular}{@{}lccc}
\hline
 & \multicolumn{3}{c}{Sample}\\
 & $z\sim1.5$ & $z\sim3$ & $z\sim4$\\
\hline
$\log(\textrm{SFR}_0\,[\msun \rmn{yr}^{-1}])$ &$-5.7\pm0.7$ &$-5.4\pm0.4$ & $-4.7\pm1.0$\\[0.1cm]
$\alpha$ &$0.70\pm0.07$ &$0.73\pm0.04$ &$0.66\pm0.10$ \\[0.1cm]
\hline
\end{tabular}
\\
The fits are performed assuming that $\textrm{SFR} = \textrm{SFR}_0M_{*}^{\alpha}$.
\end{table}

The measurements presented above yield average estimates of \lir as a
function of stellar mass at $z\sim1.5, 3,$ and 4. We can combine these
measurements with those of the observed, uncorrected UV luminosities to
obtain a total star formation rate as:

\begin{equation}
  \rm{SFR} = SFR_{\rm IR} + SFR_{\rm UV}
\end{equation}

with 
\begin{eqnarray}
  \rm SFR_{\rm IR}[\msun \rmn{yr}^{-1}] & = & 1.09\times10^{-10} L_{\rm IR} [\rm L_{\odot}]\\ 
  \rm SFR_{\rm UV}[\msun \rmn{yr}^{-1}] & = & 1.70\times10^{-10} L_{\rm FUV} [\rm L_{\odot}]
\end{eqnarray}
where we use the factors from \citet{Kennicutt_1998} that
we converted from a \citet{Salpeter_1955} to a \citet{Chabrier_2003}
IMF.  


We show in Fig. \ref{fig_stack_sfr} the average SFR-mass relations we
obtain at $z\sim1.5$, $z\sim3$, and $z\sim4$, along with best fits
from a number of previous studies (references on the figure). We find
that there are well defined average SFR-mass relations in our
UV-selected samples at the epochs we focus on. The SFR-mass relations
at $z\sim4$ and $z\sim3$ are similar to each other, while at a given
$M_*$ the SFR is around 4 times lower at $z\sim1.5$.  

We note that SFR is here equivalent to \lir for $M_* \ga
10^{10}\,\msun$, the UV contribution to the SFR being negligible, as
$L_{\rm IR}/L_{\rm FUV} >10$ in this range of masses (see
Fig. \ref{fig_stack_M*}). Fig. \ref{fig_stack_sfr} shows that
UV-selected samples do probe the ULIRGs regime at $z\sim3,4$ for $M_*
\ga10^{10}\,\msun$ as a SFR of $100\,\msun$yr$^{-1}$ correspond
roughly to $L_{\rm IR}=10^{12}\,\lsun$. This is different from what is
suggested by Figs. \ref{fig_stack_lfuv_vs_lir} and
\ref{fig_stack_lfuv}. The origin of this difference is the underlying
relations between \luvns, \lirns, and $M_*$. When stacking as a
function of $M_*$, ULIRGs are recovered in a UV selection. There are
on the other hand not recovered while stacking as a function of
\luvns, because they are mixed with other galaxies which have fainter
\lirns. This shows that \luv is not well correlated with \lir and
$M_*$.

The SFR-mass relations we observe are well described by power laws
with an average slope of 0.7; we provide fits for these relations in
Table \ref{tab_sfr_mass_fits}.  Note nevertheless that at $z\sim1.5$,
the SFR-mass relation we observe is better described by a broken power
law, with a slope of $\sim 0.85$ for $M_*<10^{10.5}\,\msun$ and a
shallower slope $\sim 0.5$ for higher masses.

We compare our results with previous determinations of the SFR-mass at
various redshifts. At $z=1$, the average relation from
\citet{Elbaz_2007}, derived from a restframe optical selection and
using $24\,\mu$m observations to constrain the amount of dust
attenuation, has a lower amplitude than ours.  Our results at $z\sim3,
4$ and $z\sim1.5$ bracket those at $z=2$ of \citet{Daddi_2007} and
\citet{Wuyts_2011b}. \citet{Daddi_2007} based their study on a $K-$band
selection and $24\,\mu$m observations, while at the same redshift
\citet{Wuyts_2011b} used optical selections and a combination of FIR
observations (including \textit{Herschel}/PACS) and SED fitting for
dust attenuation. At $z=3$, \citet{Magdis_2010} derived a SFR$-M_*$
relation for LBGs with IRAC observations, and correcting for dust
attenuation using the UV slope of the continuum. Our results at
$z\sim3$ agree with theirs at the high mass end, but have a higher
amplitude in the lower mass range we explore. On the other hand, our
measurements are in good agreement with those from \citet[][based on a
  LBG sample, and using the slope of the UV continuum to correct for
  dust attenuation]{Bouwens_2012} at $z\sim4$ in the range of masses
where they overlap, as well as if we extrapolate them at higher
masses.  In summary, the SFR$-M_*$ relations we obtain are in good
agreement with these other studies.





\subsection{Intrinsic and observed relations between dust attenuation
  and $M_*$ for UV-selected galaxies}\label{sec_irx_bias}

\begin{figure}
\includegraphics[width=\hsize]{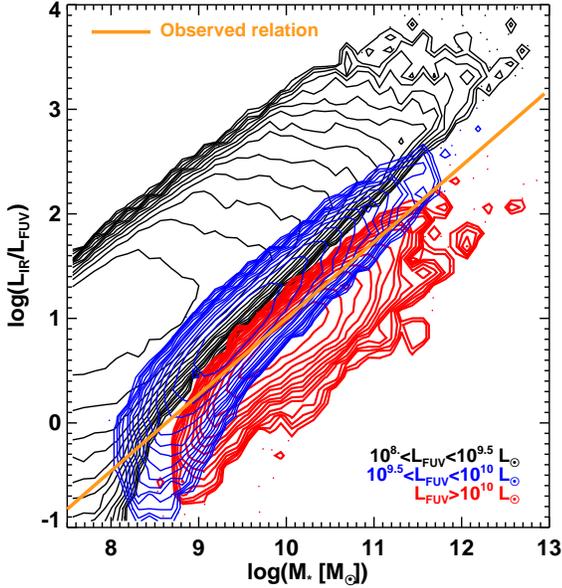}
\caption{Simulated FIR to UV luminosity ratio as a function of stellar
  mass from a mock catalogue (see text). The red ($L_{\rm
    FUV}>10^{10}\,\rm{L}_{\odot}$) and blue ($10^{9.5}<L_{\rm
    FUV}/\rm{L}_{\odot}<10^{10}$) contours show mock galaxies within
  the same range of \luv as we probe in the data. The black
  ($10^{8.}<L_{\rm FUV}/\rm{L}_{\odot}<10^{9.5}$) contours show the
  mock galaxy distribution obtained through extrapolation of the UV
  luminosity function (see text). The solid line represents the
  observed relation, determined from galaxies with $L_{\rm
    FUV}>10^{9.5}\rm{L}_{\odot}$.}
\label{fig_simul_irx}
\end{figure}


We investigate here the impact of the faint UV population on the
recovery of the relation between dust attenuation and stellar mass.
We follow the approach of \citet{Reddy_2012b} to create a mock
catalogue, which has the following properties: \luvns, \lirns, SFR,
and $M_*$. Our goal here is to model the intrinsic relation between
dust attenuation and stellar mass, by taking into account galaxies
fainter than the detection limit.

We focus here on the $z\sim1.5$ case, but show in Appendix
\ref{app_simul_irx_mass} results for $z\sim3$ and $z\sim4$. In
practice, we consider the best fit of the UV luminosity function at
$z\sim1.5$ we determined for our sample \citep{Heinis_2013}, down to
$L_{\rm FUV}=10^{8}\rm{L}_{\odot}$. We build a mock catalogue by
assigning UV luminosities according to this luminosity function. Then we
assign a FIR luminosity to each object of this catalogue. We assume that
the distribution of $\log(L_{\rm IR}/L_{\rm FUV})$ is a Gaussian. We
use as mean of this distribution the stacking results from
\citet{Heinis_2013}, and as dispersion, the dispersion required to
reproduce the few per cents of UV-selected objects detected at SPIRE
wavelength. We only have measurements for objects brighter that
$L_{\rm FUV} = 10^{9.5}\rm{L_{\odot}}$. For fainter objects, we assume
that $\log(L_{\rm IR}/L_{\rm FUV})$ is constant, as well as its
dispersion, using the results from \citet{Heinis_2013}. The values of these constants
 are $\log(L_{\rm IR}/L_{\rm FUV})_{\rmn{faint}} = 0.94$, and $\sigma(\log(L_{\rm IR}/L_{\rm
  FUV}))_{\rmn{faint}} = 0.73$. The value $\log(L_{\rm IR}/L_{\rm FUV})_{\rmn{faint}}$ is higher than 
 the average value for the sample $\langle \log(L_{\rm IR}/L_{\rm FUV})\rangle = 0.84\pm0.06$, but
  consistent with the values measured at the faint end of the sample. We determined this value in
  \citet{Heinis_2013} such that the IR luminosity function of a UV selection recovers the IR luminosity 
  function of a IR selection. Given the limited constraints on the latter, the assumption that $\log(L_{\rm 
  IR}/L_{\rm FUV})$ and its dispersion are constant for $L_{\rmn {FUV}}$ fainter that the limit of our 
  sample is necessary. The conclusions we draw from this modeling exercise would differ if the average
   IR to UV luminosity ratio for galaxies fainter than the limit of our sample is similar to that of the
    galaxies of the sample, which is unlikely given the available data.


Having now a mock catalogue with \luv and \lirns, we can assign a SFR
to each of the objects by adding the IR and UV contributions. We
finally assign a stellar mass by assuming the average SFR-mass
relation we observe at $z\sim1.5$, and assuming a dispersion of 0.15 dex
\citep{Bethermin_2012}. Note that this value might underestimate the
actual dispersion of the SFR-mass relation, but this does not have a
strong impact on our results here. We also checked that there is no
impact of incompleteness in UV on the SFR-mass relation we observe
(see Appendix \ref{app_sfr_mass}).

We show in Fig. \ref{fig_simul_irx} our modeled intrinsic IR to UV
luminosity ratio as a function of stellar mass and per bins of \luvns
from this mock catalogue. Note that we attempt to model the intrinsinc
distribution, but that our mock catalogue is also self-consistent as
we recover the observed dust attenuation-stellar mass relation for
galaxies with $L_{\rm FUV}>10^{9.5}\,\lsun$. The results from
Fig. \ref{fig_simul_irx} show that fainter objects in UV have smaller
stellar masses and higher dust attenuation. Our mock catalogue
suggests that we observe a relation between the IR to UV luminosity
ratio and $M_*$ partly because we are probing a limited range of
\luvns. We note also that we observe that the dispersion in dust
attenuation is larger for fainter galaxies \citep[see][and also
Fig. \ref{fig_stack_mstar_luv}]{Heinis_2013}. Our mock catalogue shows
that this dispersion actually originates from the $L_{\rm IR}/L_{\rm
  FUV}-M_*$ relation.

Our previous results also suggest that galaxies fainter than the
current sensitivity levels in UV restframe luminosity (i.e. down to
$u^*\sim 30$) are dustier. If that is the case, this suggests then
that the actual average relation between $L_{\rm IR}/L_{\rm FUV}$ and
stellar mass has a higher amplitude than the one we are observing, and
also that the actual dispersion in dust attenuation at a given stellar
mass is much higher, because of faint UV galaxies.

\section{Discussion}\label{sec_discussion}
\subsection{Impact of UV-selection on SFR-Mass relations}\label{sec_sel_uv_mass}
\begin{figure}
\includegraphics[width=\hsize]{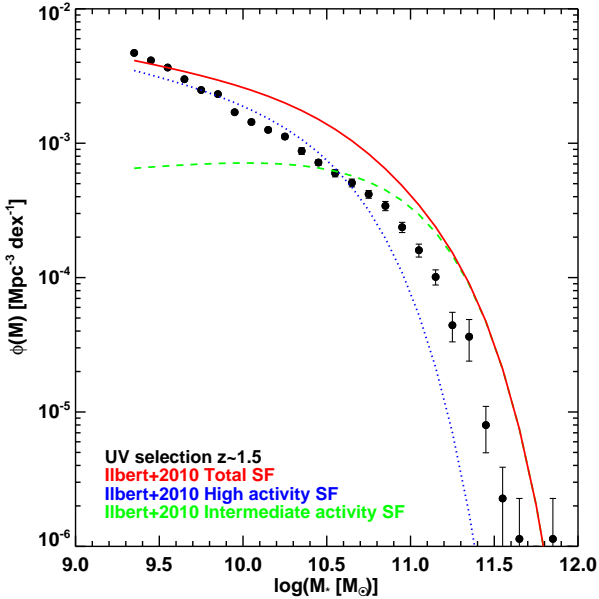}
\caption{Mass function of the $z\sim1.5$ UV-selected sample (black
  circles), compared the mass functions of star-forming galaxies in a
  $3.6\,\mu m$ selected-sample \citep{Ilbert_2010}. The dotted line
  shows the best-fit to the mass function of their high activity star
  forming galaxies, while the dashed line shows that of the
  intermediate activity star forming galaxies. The solid line shows
  the sum of these two mass functions, representing the total mass
  function of star-forming galaxies.}
\label{fig_mass_functions}
\end{figure}

We derive here average SFR-$M_*$ relations for UV-selected samples
from $z\sim4$ to $z\sim1.5$. While the relations we obtain are not
strongly sensitive to incompleteness in the UV, our results are not
drawn from a mass selection. We investigate here whether this has any
impact on our results.

We note first that we derive SFR-$M_*$ relations which have slopes
consistent with 0.7 from $z\sim4$ to $z\sim3$, which is shallower than
the value of $\sim1$ derived by a number of studies \citep{Elbaz_2007,
  Daddi_2007, Magdis_2010,Wuyts_2011b}, but in agreement with
\citet{Karim_2011, Noeske_2007, Oliver_2010, Whitaker_2012}. This
shallower slope might be caused by the fact that we are selecting
galaxies by their UV flux, and hence missing objects which have low
star formation rates.  To further examine this, we compare in
Fig. \ref{fig_mass_functions} the mass function of our sample at
$z\sim1.5$ with mass functions derived from a mass-selected sample
\citep{Ilbert_2010}, based on $3.6\,\mu$m data\footnote{The more
  recent results from \citet{Ilbert_2013} on the mass function are in
  excellent agreement with those from \citet{Ilbert_2010}; we consider
  here the earlier results as \citet{Ilbert_2010} divided their sample
  between high and intermediate activity.}. This comparison shows that
the mass function of our UV-selected sample is similar to the total
mass function of star-forming galaxies only at the low mass end, and
is otherwise lower. \citet{Ilbert_2010} also divided their sample into
high activity and intermediate activity star forming galaxies, based
on the restframe $NUV-R$ color. Fig. \ref{fig_mass_functions} shows
that the mass function of UV-selected galaxies at $z\sim1.5$ is
similar to that of high activity star-forming galaxies at
$M_*<10^{10.5}\,\msun$, while it is larger above this mass. On the
other hand, the mass function of UV-selected galaxies at $z\sim1.5$ is
lower than that of intermediate star-forming galaxies at
$M_*>10^{10.5}\,\msun$.

This comparison suggests that the UV-selection at $z\sim1.5$ is likely
to probe the full population of highly star-forming galaxies, while it
may miss roughly half the number density of intermediate star-forming
ones at $M_*>10^{10.5}\,\msun$. We note that at $z\sim3$ and $z\sim4$
UV-selected samples also miss a significant fraction of high stellar
mass galaxies. This shows that the amplitudes of our SFR-Mass
relations might be overestimated, and also that there might be an
impact on the slope of these relations, if these high stellar mass
galaxies we are missing have high SFR and large dust attenuation.


On the other hand, we
can also in this context compare our results to those from
\citet{Karim_2011}, who perform radio stacking on a mass-selected
sample. They derive SFR-mass relations which have an amplitude at most
2 times lower than ours, and a similar slope. Note that
\citet{Karim_2011} measure SFRs from stacking in VLA-radio data. While
some contamination by AGN is possible, we consider here for the
comparison their results from star-forming galaxies, which are not
expected to be dominated by radio-AGNs \citep{Hickox_2009,
  Griffith_2010}.


\subsection{Impact of star formation history on conversion from observed UV and IR luminosities to SFR}


The values of the factors commonly used to convert from UV or IR
luminosities to SFR \citep{Kennicutt_1998} assume that the star
formation has been constant over timescales of around 100 Myrs. While
useful, this assumption is not correct for galaxies with other star
formation histories. The impact of the star formation history on the
conversion from \luv or \lir to SFR has been studied by various
authors \citep[including][]{Kobayashi_2013,
  Reddy_2012b,Schaerer_2013}: in the early phases of star formation
($t<10\,$Myr), the actual conversion factors are larger than the
\citet{Kennicutt_1998} values (implying that the SFR values are
underestimated when adopting the conversion factor from \citet{Kennicutt_1998}), while for
later phases there are lower. The amplitude
of the difference depends on the star formation history, with faster
evolutions yielding larger differences. In our case, if we assume that
our SFR values are overestimated, this means that the bulk of our
samples is a population of galaxies in later phases of star formation,
with rapidly declining star formation histories, like starbursts for
instance. It is beyond the scope of this paper to characterise
precisely the star formation histories of the galaxies in our
samples. We can however base our argumentation on the results of SED
fitting of dropouts at $3<z<6$ from \citet{Schaerer_2013}. They found
that the currently available data is suggesting that these galaxies
experienced either exponentially declining or delayed star formation
histories. They also note in particular that, assuming their SED
fitting, the SFR would be slighty \textit{underestimated} if the
\citet{Kennicutt_1998} conversion factors would have been used. Moreover, \citet{Wuyts_2011a} showed by backtracing galaxies using different star formation histories that the declining star formation scenario does not enable to reproduce the number densities of star-forming galaxies between $z=4$ and $z=0$. 
In summary, given the state-of-the art SED fitting, we believe that the impact of star formation histories different from that assumed by
\citet{Kennicutt_1998} is negligible on our results.

\subsection{Evolution of specific star formation rate with redshift}
\begin{figure*}
\includegraphics[width=\hsize]{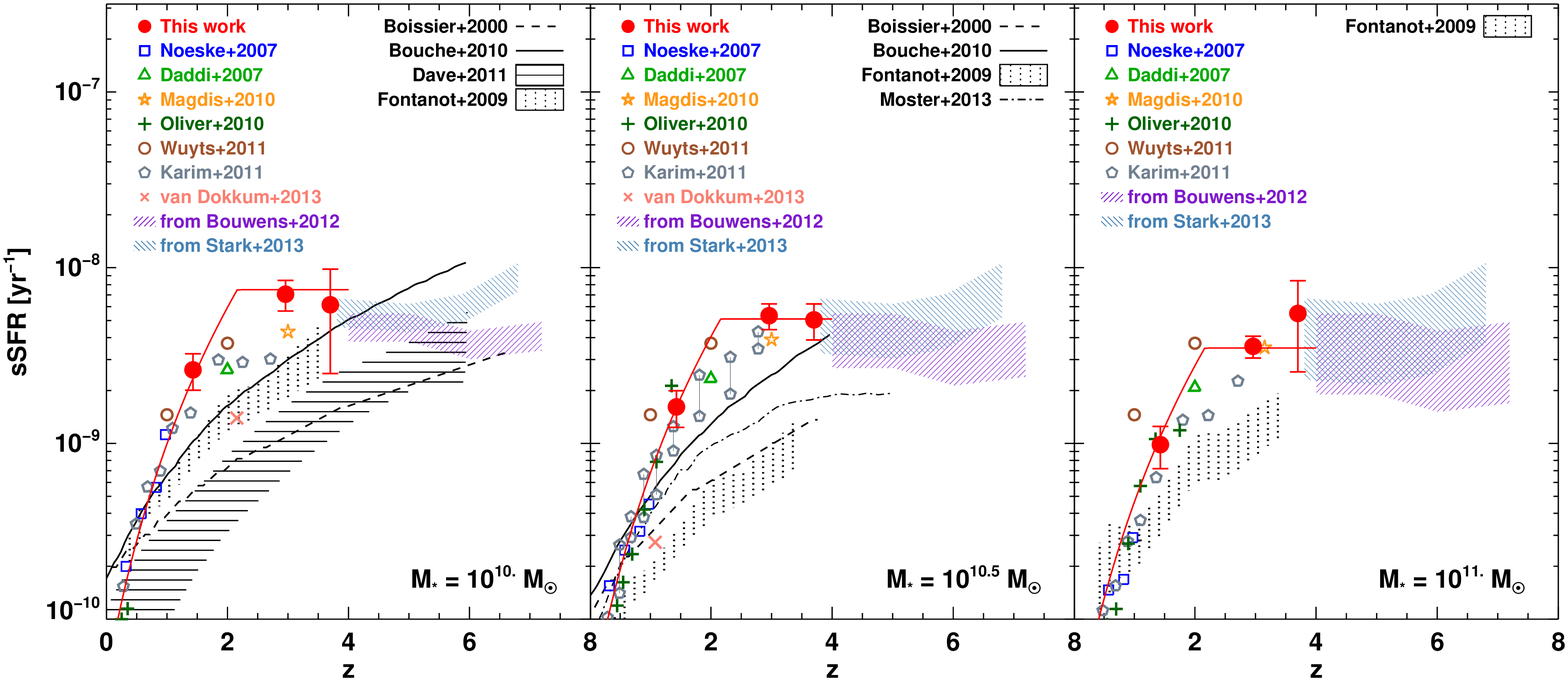}
\caption{Specific star formation versus redshift for a stellar mass of
  $10^{10}\,\msun$ (left), $10^{10.5}\,\msun$ (middle),
  $10^{11}\,\msun$ (right). Our results are showed as red filled
  circles. The red thick line is a fit to our measures using
  eq. \ref{eq_ssfr_z} (see Sect. \ref{sec_sfh_ms}). We compare our
  resuls to those of \citet[][crosses, for Milky Way-like
    galaxies]{VanDokkum_2013}, \citet[][open squares]{Noeske_2007},
  \citet[][open triangle]{Daddi_2007}, \citet[][open star, shifted by
    $+0.15$ in redshift at $M_* = 10^{11}\,\msun$ for
    clarity]{Magdis_2010}, \citet[][plus symbols]{Oliver_2010},
  \citet[][open circles]{Wuyts_2011b}, \citet[][star forming sample,
    open pentagons]{Karim_2011}, and to the range of values allowed by
  the results of \citet[][diagonally hatched regions, see text for
    details]{Bouwens_2012, Stark_2013}. At $M_*=10^{10.5}\,\msun$, for the values
  of \citet{Karim_2011}, we show their results at
  $M_*\sim10^{10.37}\,\msun$ and $M_*\sim10^{10.75}\,\msun$, as they
  do not list measurements at $M_*=10^{10.5}\,\msun$. We compare these
  observations to the models of \citet[][dashed line]{Boissier_2000},
  \citet[][solid line]{Bouche_2010}, \citet[][horizontally hatched
    regions]{Dave_2011}, \citet[][vertically hatched
    regions]{Fontanot_2009} and \citet[][dot dashed
    line]{Moster_2013}.}
\label{fig_ssfr_z}
\end{figure*}

Our measurements show that the amplitude of the SFR-$M_*$ relation is
similar between $z\sim4$ and $z\sim3$, and then decreases significantly from
$z\sim3$ to $z\sim1.5$. Another way to look at these results is to consider
the specific star formation rate $\rm{sSFR} = \rm{SFR}/M_*$ which is
an indicator of star formation history, in the sense that it is the
inverse of the time needed for a galaxy to double its mass if it has a
constant SFR.

We show in Fig. \ref{fig_ssfr_z} the evolution with redshift of the
specific star formation rate for three mass bins: $10^{10}\,\msun$,
$10^{10.5}\,\msun$, and $10^{11}\,\msun$. We compute the average SFR
for our samples by stacking galaxies in bins of stellar mass centered
on these values, with sizes of 0.2\,dex at $z\sim1.5$ and $z\sim3$, and a
size of 0.4\,dex at $z\sim4$.

We compare our results to the measurements of \citet{Daddi_2007,
  VanDokkum_2013, Karim_2011, Magdis_2010, Noeske_2007,
  Wuyts_2011b}. At $z>4$, there are basically no results yet in the
mass range we explore. We show here an extrapolation of the results
from \citet{Bouwens_2012} and \citet{Stark_2013}. \citet{Bouwens_2012} give values of sSFR at $M_* =
5\times10^9\,\msun$ corrected from dust attenuation (based on the UV
slope of the continuum), using their own sample at $z=4$, and the
results from \citet{Stark_2009} and \citet{Gonzalez_2010} at higher
redshifts. \citet{Stark_2013} derive sSFRs also at $M_* = 5\times10^9\,\msun$ at $4<z<7$, taking into account the impact of emission lines on the measure of stellar masses, and correcting from dust attenuation using the slope of the UV continuum.
We extrapolate results from both studies in our mass range assuming
that there is a power law relation between SFR and stellar mass at
$z>4$, and that the slope of this relation is between 0.7 (the value
measured at $z=4$ by \citet{Bouwens_2012}, also consistent with our
results) and 1 \citep[closer to the value observed at lower redshifts
by other studies like][]{Wuyts_2011b}.

Our results are in overall agreement with previous measurements at
$z\sim1.5$. Note that all measurements are significantly higher than
those of \citet{VanDokkum_2013}, who derived the star formation
history of Milky Way-like galaxies (see Sect. \ref{sec_sfh_ms} for
further discussion).

 At $z\sim3$, our measurements are quite high compared to
the values from previous studies, in particular at
$M_*=10^{10}\,\msun$. In this mass bin, our estimates are larger than
the measurements from \citet{Karim_2011} and \citet{Magdis_2010}, but
they are consistent at $1.2\,\sigma$ and $0.3\,\sigma $
respectively. In other word, our sSFR results represent the upper
range of available measurements. Note however that our results are in
very good agreement with those of \citet{Magdis_2010} at $z\sim3$ for
$M_*=10^{10.5.}\,\msun$ and $M_*=10^{11.}\,\msun$.

At $z\sim4$, our results agree with those from \citet{Bouwens_2012}  and \citet{Stark_2013} at
$z=4$. Our results are also in agreement with the sSFR being constant at $3<z<4$, while the results of \citet{Stark_2013} suggest that the sSFR is increasing at higher redshifts ($z>5$).

We compare our results with a few models, from \citet{Boissier_2000},
\citet{Bouche_2010}, \citet{Dave_2011}, \citet{Fontanot_2009}, and
\citet{Moster_2013}. These models are quite different and give a
sample of various simulation techniques available. We briefly describe
all of them.


\citet[][see also \citet{Boissier_1999} and
\citet{Boissier_2010}]{Boissier_2000} built an analytical model which
predicts the chemical and spectrophotometric evolution of spiral
galaxies over the Hubble time. This model reproduces a large number of
present properties of the Milky Way and local spiral galaxies (such
as: color-magnitude diagrams, luminosity-metallicity relationship, gas
fractions, as well as color and metallicity
gradients). \citet{Bouche_2010} based their model under the assumption
that the gas accretion in galaxies is mostly driven by the growth of
dark matter haloes \citep[e.g.][]{Dekel_2009}. They also assume that
the gas accretion efficiency decreases with cosmic time, and is only
efficient for dark matter haloes of masses
$10^{11}<M_h/\msun<1.5\times10^{12}$. \citet{Dave_2011} ran
hydrodynamical simulations which include galactic outflows,
implementing several models for winds; we show on
Fig. \ref{fig_ssfr_z} the range of sSFR spanned by these models,
including the model without winds. \citet{Fontanot_2009} compared the
predictions from three semi-analytical models, namely those of
\citet{DeLucia_2007, Monaco_2007, Somerville_2008}. All three models
are based on the combination of dark matter simulations complemented
by empirical relations for baryonic physics. All these models include
supernovae and AGN feedback. We show on Fig. \ref{fig_ssfr_z} the
range of sSFR spanned by these three models. \citet{Moster_2013} studied the
mass assembly of galaxies using abundance matching models, by matching
observed stellar mass functions simultaneously at various redshifts.

The comparison in Fig. \ref{fig_ssfr_z} of observations and models
shows that models match the observations roughly well at low redshift \citep[$z<0.5$, 
see also e.g.][]{Damen_2009}, underestimate the sSFR up to $z=4$, and are 
potentially in better agreement at higher redshifts. An interesting point is that the models
we consider here are quite different in terms of implementation and
assumptions; however they all predict a similar evolution which does
not match the observations for $0.5\la z\la4$. At $M_*=10^{10}\msun$, the model of
\citet{Bouche_2010} and the compilation of models from
\citet{Fontanot_2009} are the closest to the observations among the
ones we consider here. Still, these models do not reproduce the high
sSFR we observe at $z\sim3$. At $M_*=10^{10.5}\msun$, the model of
\citet{Bouche_2010} presents the same level of agreement with our
measurements, while the discrepancies between the compilation of
\citet{Fontanot_2009} and the observations are more important. We note
also that all these models are actually more or less consistent with
the redshift evolution expected according to the cold gas accretion
scenario \citep{Dekel_2009}. This scenario predicts that the baryonic
accretion onto galaxies follows directly the dark matter accretion
onto dark matter haloes, and evolves as $\dot{M} \propto
(1+z)^{2.25}$. Our results show that this scenario is in agreement
with the observations for $0< z \la 1.5$, but is less efficient at
reproducing galaxies properties at $1.5\la z\la3$.


There has been some attempts to reconcile model predictions with the
observations of the redshift evolution of the sSFR. \citet{Dave_2008}
noted that a number of observations suggest that the IMF is not
universal and could evolve with redshift, in the sense that it would
be weighted towards more massive stars at high redshift. Such an IMF
would imply that SFRs as derived here are \textit{overestimated} with
respect to using an evolving IMF, by a factor that increases with
redshift, being around 4 at $z=4$. Whether the IMF is universal, or evolves with
redshift, remains to date a controversial subject. Indeed recent studies suggest in contrary to
 \citet{Dave_2008} that there is observational evidence for bottom-heavy IMF at high redshift \citep[see e.g.][]
{VanDokkum_2012}. 

\citet{Weinmann_2011} considered a
number of modifications to semi-analytical models in order to match
the observed redshift evolution of the sSFR. They found that models
can match the observations at $z>4$ if there is either strong stellar
feedback at high redshift at all masses, or inefficient star
formation. At $z=2-3$, where the models underpredict the sSFR, the
feedback could drop, or gas which was prevented to form stars earlier
could be at that time available for star formation.  We provide new
and improved observational constraints to test these scenarios. Future
observations of the gas content of high redshift galaxies will also
enable to discriminate between those.

\subsection{The star formation histories of Main sequence
  galaxies}\label{sec_sfh_ms}


 \begin{figure}
\includegraphics[scale=0.45]{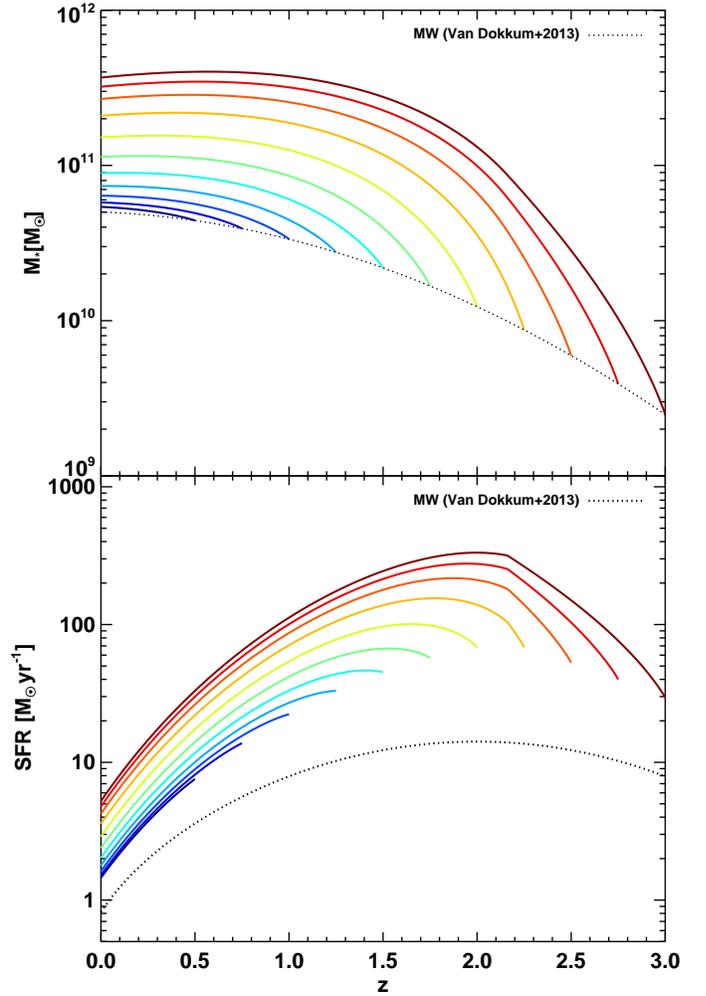}
\caption{Star formation history of Main Sequence galaxies which have
  the same stellar mass as the Milky Way. We integrate here the sSFR
  of the Main Sequence; we use the best fit to our measurements of
  $\rmn{sSFR_{\rmn{MS}}}(z,M_*)$ (eq. \ref{eq_ssfr_z}). \textit{Top:}
  evolution of the stellar mass for Main Sequence galaxies, assuming
  that galaxies have the same stellar mass as the Milky Way at 10
  equally spaced redshifts values ranging from $z=3$ to $z=0.5$
  according to the measurements of \citet{VanDokkum_2013} (dotted
  line); the redshifts are color-coded: bluer for lower initial
  redshift for the integration boundary condition. \textit{Bottom:}
  Same as top plot, but for the star formation rate, with
  corresponding colors.}
\label{fig_sfr_mass_integration}
\end{figure}

Our measurements bring new constraints at high redshift on the sSFR of
the Main Sequence galaxies. We can use these results to derive the
star formation history of galaxies staying on the Main Sequence. We
first recall that galaxies can not remain on the Main Sequence from
high redshift to $z=0$, given the stellar masses and SFR they would
have in the local Universe.  We then give estimates of the timescale
galaxies can stay on the Main Sequence before quenching of the star
formation.

We consider here a parameterised form of the dependence with redshift
and stellar mass of the sSFR of the Main Sequence. We follow the
approach of \citet{Bethermin_2012}, and we assume that:
\begin{eqnarray}\label{eq_ssfr_z}
\rmn{sSFR}_{\rmn{MS}}(z, M_*) & = & \rmn{sSFR_{\rmn{MS, 0}}}\times \left(\frac{M_{*}}{10^{11}\,\msun}\right)^{\beta_{\rmn{MS}}}\\ \nonumber
& & \times\left(1+\min(z, z_{\rm evo})\right)^{\gamma_{\rmn{MS}}}\rmn{,}
\end{eqnarray}

where $\rmn{sSFR_{\rmn{MS, 0}}}$ is the sSFR of the Main Sequence at
$z=0$ for galaxies of $M_* = 10^{11}\msun$, $\beta_{\rmn{MS}}$ is the
slope of the sSFR-$M_*$ relation, and $\gamma_{\rmn{MS}}$ encodes the
power-law redshift evolution of the amplitude of the sSFR-$M_*$
relation. We modify the values of these parameters to match our
measurements as well as the measurements at lower redshifts from
\citet{Noeske_2007}: $\rmn{sSFR}_{\rmn{MS, 0}}=10^{-10.66}
\rmn{yr}^{-1}$, $\beta_{\rm MS}=-0.33$, $z_{\rm evo} = 2.16$, and
$\gamma_{\rm MS}=4.4$. We show the resulting sSFR evolution using
these parameters as a red line on Fig. \ref{fig_ssfr_z}.

We note that eq. \ref{eq_ssfr_z} can also be written as
\begin{eqnarray}\label{eq_ssfr_t}
  \frac{1}{1-R}\frac{1}{M_*}\frac{\rmn{d}M_*}{\rmn{d}t} & = & \rmn{sSFR_{MS, 0}}\times \left(\frac{M_{*}}{10^{11}\,\msun}\right)^{\beta_{\rm MS}}\\ \nonumber
& & \times\left(1+\min\left(z(t), z(t_{\rm evo})\right)\right)^{\gamma_{\rm MS}}
\end{eqnarray}

where $t_{\rm evo}$ is the lookback time corresponding to $z_{\rm
  evo}$. We wrote the SFR in terms of the derivative of $M_*$ with
respect to time assuming that

\begin{equation}
  \frac{\rmn{d}M_*}{\rmn{d}t} = \rmn{SFR}(1-R)\rmn{.}
\end{equation}
$R$ is the return fraction, that we set to \citep{Conroy_2009}
\begin{equation}
  R = 0.05\ln\left(1+\frac{\Delta t}{0.03\rmn{Myr}}\right)
\end{equation}
where $\Delta t$ is the time elapsed since the formation of stars.

We can then use the fact that eq. \ref{eq_ssfr_t} is a differential
equation for $M_*(z)$. We obtain $M_*(z)$, and from this
SFR$(z)$. This procedure requires boundary conditions of stellar mass
at a given redshift. In other words, we can start the integration of
eq. \ref{eq_ssfr_t} at any redshift, but we need to choose an initial
stellar mass at this redshift. This means that we are making galaxies 'enter' on the Main Sequence at these stellar mass and redshift. We are considering here only the mean location of the Main Sequence. This means that, prior to entering the Main Sequence in the sense of this simple model, galaxies could for instance be lower in the SFR-Mass plane, but still within the Main
 sequence at redshifts higher than this initial redshift.

We consider here the result of \citet{VanDokkum_2013}, who derive the star formation history of Milky
Way-like galaxies, by studying up to $z=2.5$ galaxies with the same
number density as galaxies with the stellar mass of Milky Way at
$z=0$. \citet{VanDokkum_2013} derive the redshift evolution of the
stellar mass of such galaxies. We use their fit to get initial stellar
mass at a given redshift\footnote{\citet{VanDokkum_2013} discuss that
  major mergers are not expected to play a significant role in the
  star formation history of Milky Way-like galaxies.}.


We integrate eq. \ref{eq_ssfr_t} down to $z=0$, starting from various
initial redshifts, which we consider between $z=3$ and $z=0.5$. We
show on Fig. \ref{fig_sfr_mass_integration} the evolution of the
stellar mass and SFR for galaxies which remain on the Main Sequence
and have the same stellar mass as the Milky Way at these initial
redshifts. Doing so we look at the star formation history of galaxies
which have the same stellar mass as the Milky Way at these initial
redshifts, and stay on the Main Sequence until $z=0$\footnote{We
  assumed here that eq. \ref{eq_ssfr_t} is valid at all stellar
  masses. It has been suggested that the relation between the sSFR and
  $M_*$ flattens below a given mass, which might evolve with redshift
  \citep[`crossing mass',][]{Karim_2011}. We checked that including a
  flattening of the sSFR at low masses does not have a strong impact
  on our conclusions here.}.

Assuming that a galaxy is on the Main Sequence for $1\la z<3$ leads to
much higher SFR and stellar mass than the Milky Way at $z=0$. On the
other hand, if we assume that the Milky Way is on the Main Sequence
between $z=0.5$ and $z=0$, we obtain a stellar mass similar to the
Milky Way at $z=0$, and a SFR around 2 times higher. Note that
galaxies with $M_* \sim 10^{10}\,\msun$ at $z\sim 2$ would have $M_*
\sim 2\times10^{11}\,\msun$ at $z=0$. This is in strong disagreement
with measurements of the redshift evolution of the stellar mass
functions of star forming galaxies \citep[e.g.][]{Ilbert_2010} which
show little evolution between $z=2$ and $z=0$ at the high mass end. The star formation
histories on Fig. \ref{fig_sfr_mass_integration} are actually quite
different from that expected for the Milky Way (dotted line on bottom
panel), even though we assumed the observed stellar mass of Milky
Way-like galaxies at various redshifts as boundary conditions. This is
actually due to the fact that the Milky Way is not on the mean
location of the Main Sequence for $1<z<2$ (see crosses showing the
measurements of \citet{VanDokkum_2013} on
Fig. \ref{fig_ssfr_z}). Assuming the values from
\citet{VanDokkum_2013} and the results of \citet{Wuyts_2011b} for the
distribution of galaxies in the $(\rmn{SFR},M_*)$ plane suggests that
the Milky Way is rather on the lower enveloppe of the Main Sequence
for $0<z<2$.  Our results suggest on the other hand that the sSFR of
star-forming galaxies is quite high at $z=3,4$, which yields a high
SFR peak in the derived star formation histories.

The results shown on Fig. \ref{fig_sfr_mass_integration} suggests that
the assumption that galaxies remain on the Main Sequence until $z=0$
is not correct. The consequence is that the Main Sequence is built of
different star-forming galaxies at various redshifts.


\begin{figure}
\includegraphics[width=\hsize]{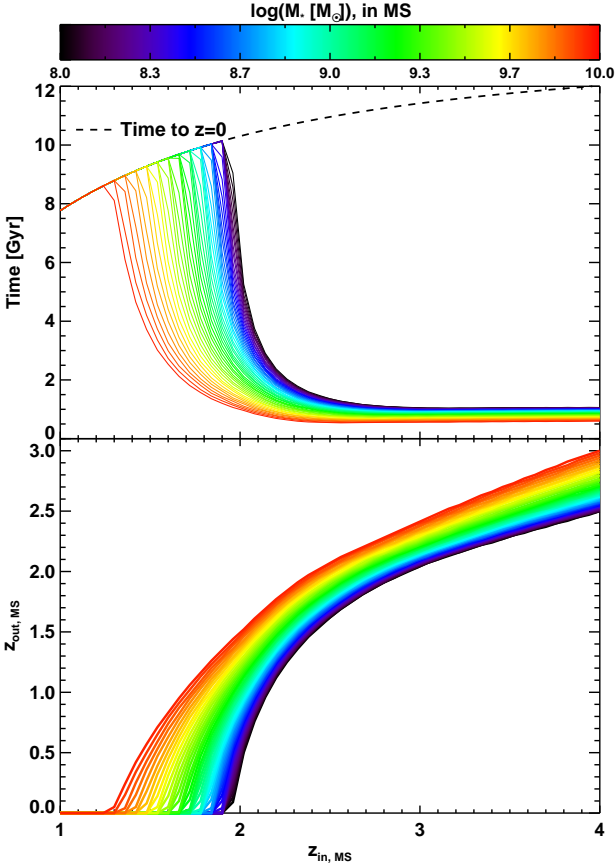}
\caption{\textit{Top}: time galaxies stay on the Main Sequence before
  quenching of star formation, as a function of the redshift they
  enter on the Main Sequence.  The dashed line shows the time left
  until $z=0$. \textit{Bottom}: redshift of galaxies when they exit
  the Main Sequence, as a function of the redshift they enter on the
  Main Sequence. In both plots, the color codes the stellar mass of
  galaxies at the time they enter on the Main Sequence, ranging from
  $10^8$ to $10^{10}\,\msun$.}
\label{fig_time_MS}
\end{figure}

These results raise the question of the amount of time galaxies can
stay on the Main Sequence. In order to determine this time, we need to
define a criterion to determine the epoch when galaxies exit the Main
Sequence. We use here the `quenching mass' ($M_{\rm Q}$) as defined by
\citet{Ilbert_2013}. We used the same method as above to investigate
this. We consider once again eq. \ref{eq_ssfr_t}, but this time we
stop the integration, i.e. we make galaxies exit the Main Sequence, at
the redshift when their stellar mass is larger than the quenching mass
at the same time. Galaxies experiencing quenching of star formation
exit the Main Sequence by going down in the $(\rmn{SFR},M_*)$ plane at
a given M$_*$ \citep[e.g.][]{Wuyts_2011b}. We do not consider here
starbursts galaxies as they represent a significantly smaller number
density \citep{Rodighiero_2011}.

We follow \citet{Ilbert_2013} and assume that the quenching mass is
the mass where the number density of quiescent galaxies is maximum. We
consider the measurements from \citet{Ilbert_2013} of the mass
function of quiescent galaxies (available for $0.5<z<3$) and
complement them at $z=0$ by the measurement of
\citet{Baldry_2012}. The evolution with redshift of the quenching mass
can be adjusted to the following form:


\begin{equation}
  M_{\rm Q}(z)[\msun] = 3.7\times10^{10}\times(1+z)^{0.53}
\end{equation}




We make the galaxies enter the Main Sequence at redshifts $1<z<4$, and
at masses in the range $10^8<M_*/\msun<10^{10}$. We show the time
galaxies stay on the Main Sequence in Fig. \ref{fig_time_MS}. We perform the
 integration only until $z=0$; in other words, we do not derive times larger than the time to $z=0$ for
 galaxies that have not reached $M_{\rm Q}$ at $z=0$. This means that galaxies that are still on the Main
 Sequence at $z=0$ are represented by locations on the dashed line on the top panel of Fig.
  \ref{fig_time_MS}, or at $z=0$ in the bottom panel.

Given our assumptions, our results show that galaxies which enter the
Main Sequence at $z<4$ stay on it at least 1\,Gyr. As expected, at a
given entrance redshift on the Main Sequence, less massive galaxies
spend more time on the Main Sequence to reach the quenching
mass. Galaxies entering on the Main Sequence at $2.5<z<4$ stay around
1\,Gyr on it. At lower redshifts, the quenching mass decreases, but
the average sSFR also decreases, which in turn yields that galaxies
stay longer on the Main Sequence. For instance, with the scenario we
consider here, galaxies with masses $10^8<M_*/\msun<10^{10}$ which
enter the Main Sequence at $z<1.2$ stay on the Main Sequence until
$z=0$. \citet{Leitner_2012} and \citet{Zahid_2012} reach similar
conclusions regarding the star formation histories of Main Sequence
galaxies at $z<2$.

We assumed here that the sSFR is constant for $z>2.16$. Assuming that the sSFR
 increases with $z$ from $z\sim3$ \citep[see e.g.][]{Stark_2013} would mean faster evolution for high
 redshift galaxies, implying: stronger disagreement for the evolution of the Milky Way as discussed
here, and shorter times on the Main sequence for high redshift galaxies.
We note that the simplistic calculation presented here requires to be tested against
the redshift evolution of the stellar mass functions of quiescent and
star forming galaxies, which is beyond the scope of this paper, and
will be the subject of forthcoming work.

\section{Conclusions}\label{sec_conclusion}
We studied the FIR properties of large samples of UV-selected galaxies
at $1.5<z<4$, by combining the COSMOS multiwavelength dataset with the
HerMES/\textit{Herschel} SPIRE imaging. We measured by stacking the
average IR luminosity as a function of UV luminosity, stellar mass,
and both. Our results can be summarised as follows:
\begin{enumerate}[1.]
  \item At $z\sim1.5$, there is a good correlation between \lir and
    \luvns ($8\times10^{9}<L_{\rm FUV}/L_{\odot}<5\times10^{10}$), while at $z\sim3$ and $z\sim4$, \lir and \luv are not well
    correlated.
    \item Consequently, the ratio \lir$/$\luv at $z\sim3,4$ is
      decreasing with \luvns.
    \item The average dust
      attenuation (as traced by the \lir$/$\luv ratio) is well
      correlated with stellar mass at $1.5<z<4$, and does not show
      significant evolution in this redshift range, in the range of
      masses we explore.
  \item We investigated the joint dependence of dust attenuation with
    stellar mass and \luvns. While well correlated with stellar mass, dust attenuation also shows secondary
    dependence on \luvns. At a given stellar mass, dust attenuation
    decreases with \luvns; at a given \luvns, dust attenuation
    increases with stellar mass. We also provide empirical relations
    between dust attenuation, $M_*$, and \luvns, at $z\sim1.5$ and $z\sim3$.
  \item The average SFR-$M_*$ relations for UV-selected samples at
    $1.5<z<4$ are well approximated by a power law, with a slope of around 0.7. At a given stellar
    mass, the average SFR is similar at $z\sim3$ and $z\sim4$, but is 4
    times higher than at $z\sim1.5$.
  \item Our results provide new constraints on the sSFR at
    $1.5<z<4$. Current models of galaxy formation and evolution do not
    reproduce accurately the sSFR evolution we observe, in particular
    at $z\sim3$ and $z\sim4$, where standard models underpredict the
    observations.
  \item We use our results for the evolution of the sSFR with redshift
    to characterise the star formation histories of Main Sequence
    galaxies. We find that galaxies would have too large stellar
    masses if they stay on the Main Sequence from high redshift to
    $z=0$. Assuming that galaxies exit the Main Sequence when their
    stellar mass is equal to the `quenching mass', we determine the
    time galaxies stay on the Main Sequence. This suggests that
    galaxies stay around 1\,Gyr on the Main Sequence at high redshift
    ($2.5<z<4$), while they stay longer on the Main Sequence at lower
    redshifts. For instance, Main Sequence galaxies (with
    $10^8<M_*/\msun<10^{10}$) at $z=1$ stay until $z=0$ on the Main
    Sequence, as they do not reach the quenching mass.
\end{enumerate}

\section*{Acknowledgments}
We thank the referee for useful comments and suggestions. S.H. and V.B. acknowledge support from the Centre National d'Etudes Spatiales. We thank the COSMOS team for sharing data essential to
this study. SPIRE has been developed by a consortium of institutes led
by Cardiff Univ. (UK) and including: Univ. Lethbridge (Canada); NAOC
(China); CEA, LAM (France); IFSI, Univ. Padua (Italy); IAC (Spain);
Stockholm Observatory (Sweden); Imperial College London, RAL,
UCL-MSSL, UKATC, Univ. Sussex (UK); and Caltech, JPL, NHSC,
Univ. Colorado (USA). This development has been supported by national
funding agencies: CSA (Canada); NAOC (China); CEA, CNES, CNRS
(France); ASI (Italy); MCINN (Spain); SNSB (Sweden); STFC, UKSA (UK);
and NASA (USA). The data presented in this paper will be released
through the {\em Herschel} Database in Marseille HeDaM
(\url{http://hedam.oamp.fr/HerMES})

\appendix

\section{Empirical recipes for dust attenuation correction}\label{app_emp_irx}
\begin{figure}
\includegraphics[width=\hsize]{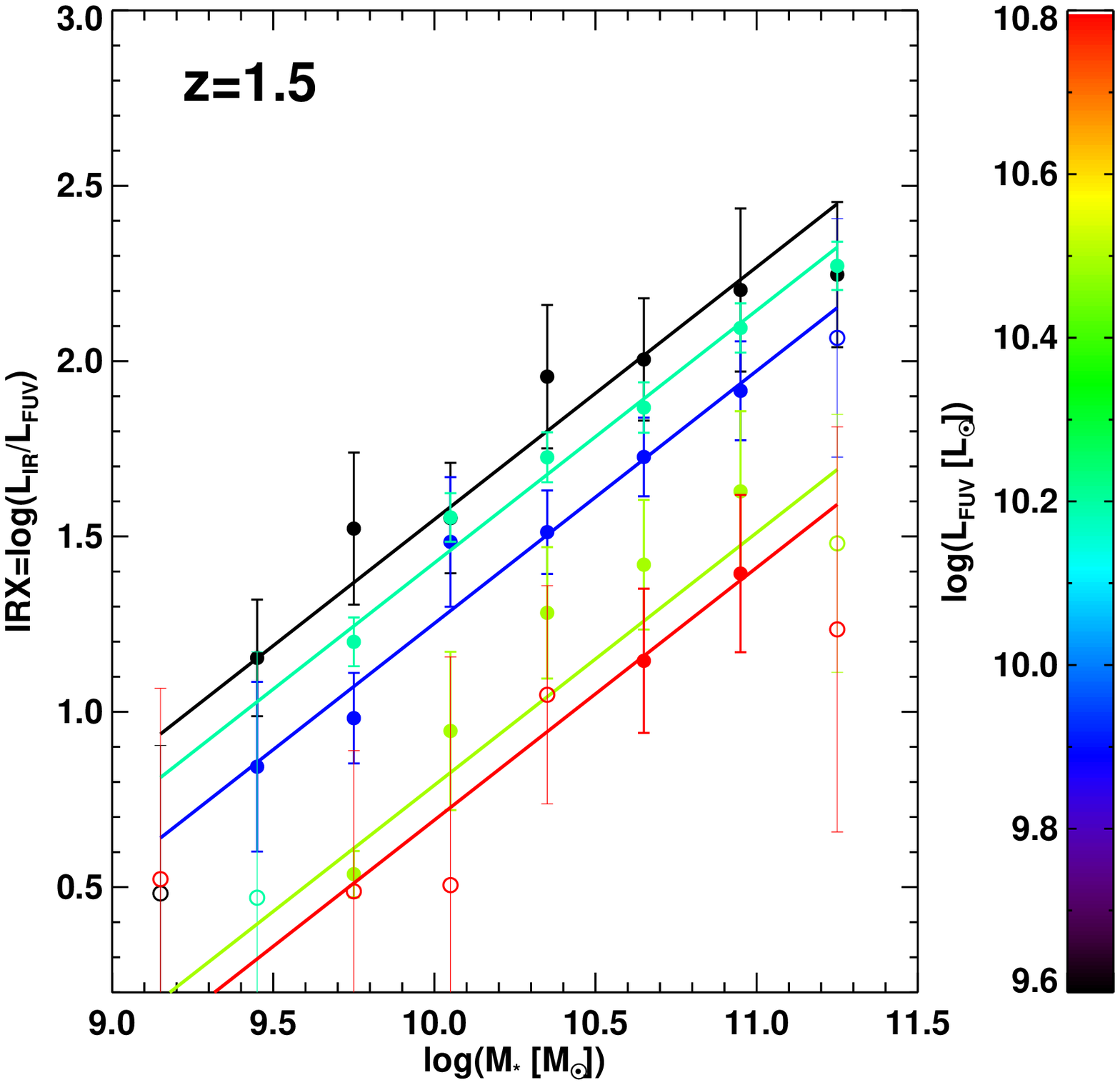}
\caption{Infrared to UV luminosity ratio as a function of stellar
  mass, at $z\sim1.5$. We show here the results from
  Fig. \ref{fig_stack_mstar_luv} along with fits by power laws. The
  mean UV luminosity is color-coded. Filled symbols represent stacking
  measurements with $S/N >3$ in all SPIRE bands, and open symbols other
  stacking measurements.}
\label{fig_irx_luv_mstar_fit}
\end{figure}

\begin{figure}
\includegraphics[width=\hsize]{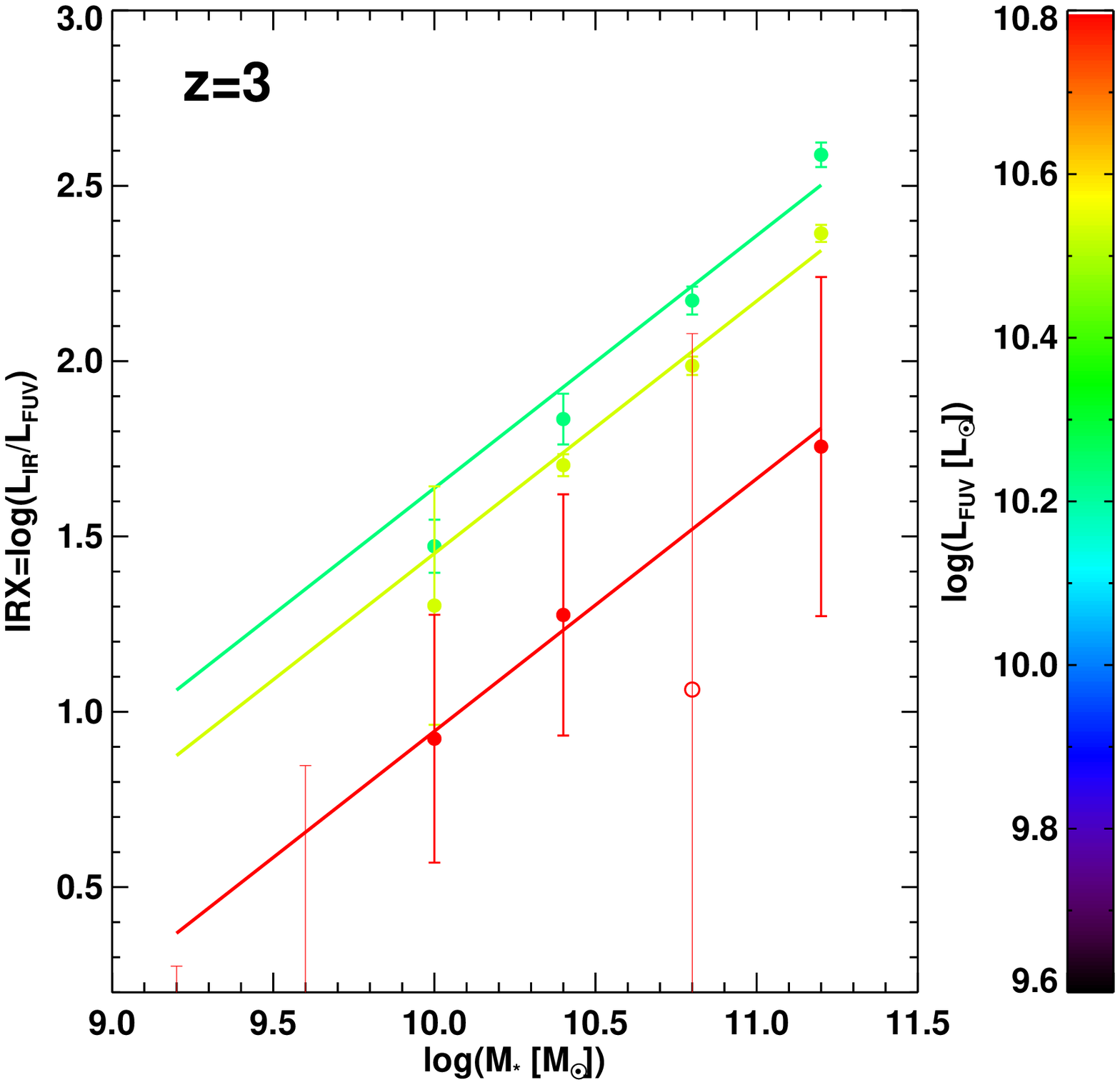}
\caption{Same as Fig. \ref{fig_irx_luv_mstar_fit}, for the $z\sim3$
  sample.}
\label{fig_irx_luv_mstar_fit_z3}
\end{figure}

We provide here empirical relations to correct for dust attenuation,
given observed UV luminosity and stellar mass. We show the relations
between the infrared to UV luminosity ratio and the stellar mass, for
several bins of UV luminosity, at $z\sim1.5$
(Fig. \ref{fig_irx_luv_mstar_fit}) and at $z\sim3$
(Fig. \ref{fig_irx_luv_mstar_fit_z3}). These measurements are the same
as those presented in Fig. \ref{fig_stack_mstar_luv}. We assume that
\begin{equation}
  \rmn{IRX}(L_{\rm FUV}, M_*) = \rmn{IRX_0}(L_{\rm FUV}) + \delta(L_{\rm FUV})\log\left(\frac{M_{*}}{10^{10.35}}\right)
\end{equation}

where we set here $\delta(L_{FUV}) = 0.72$, which is the slope of the
IRX$-M_*$ correlation for the full sample at $z\sim1.5$, and is also
valid at $z\sim3$. We provide the best fit values for IRX$_0(L_{\rm FUV})$
in Table \ref{tab_fit_irx_mstar_lfuv_high_sn_z1p5} for $z\sim1.5$
measurements and in Table \ref{tab_fit_irx_mstar_lfuv_high_sn_z3} for
$z\sim3$ measurements. We include only the stacking measurements with $S/N>3$ in
the fit, but including other stacking measurements does not have an impact
on the results.

\begin{table}
\caption{$z\sim 1.5$ best fit of IRX$-M_*-L_{\rm FUV}$ relation for stacking measures with $S/N >3$ in all SPIRE bands}
\label{tab_fit_irx_mstar_lfuv_high_sn_z1p5}
\begin{tabular}{@{}cc}
\hline
$\log(L_{\rm FUV} [L_{\odot}])$ range & IRX$_0$ \\
\hline
9.44 -- 9.74 &  1.80 $\pm$ 0.07\\
9.74 -- 10.04 &  1.50 $\pm$ 0.06\\
10.04 -- 10.34 & 1.68 $\pm$ 0.03\\
10.34 -- 10.64 &  1.04 $\pm$ 0.05\\
10.64 -- 10.94 &  0.94 $\pm$ 0.15\\
\hline
\end{tabular}
\medskip
\end{table}

\begin{table}
\caption{$z\sim3$ best fit of IRX$-M_*-L_{\rm FUV}$ relation for stacking measures with $S/N >3$ in all SPIRE bands}
\label{tab_fit_irx_mstar_lfuv_high_sn_z3}
\begin{tabular}{@{}cc}
\hline
$\log(L_{\rm FUV} [L_{\odot}])$ range & IRX$_0$ \\
\hline
10.09 -- 10.39 & 1.89 $\pm$ 0.02\\
10.39 -- 10.69 & 1.70 $\pm$ 0.01\\
10.69 -- 10.99 & 1.20 $\pm$ 0.21\\
\hline
\end{tabular}
\medskip
\end{table}

\section{Impact of UV incompleteness on SFR-Mass relation}\label{app_sfr_mass}
\begin{figure}
\includegraphics[width=\hsize]{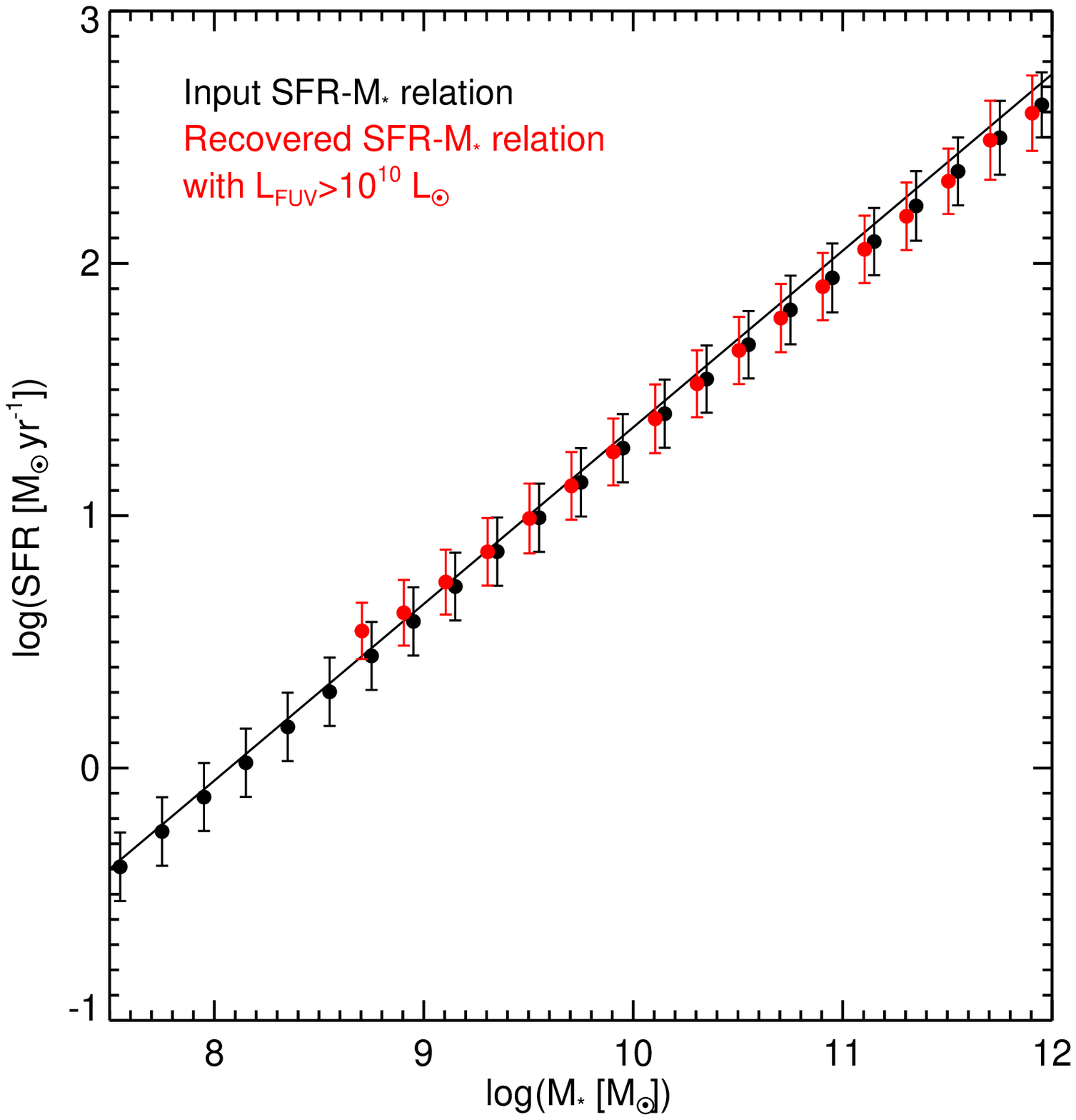}
\caption{Simulated SFR-Mass relation at $z\sim1.5$ from a mock
  catalogue. The black circles show the input SFR-Mass relation, and the
  red circles the recovered relation for galaxies brighter than
  $L_{\rm FUV}=10^{10}\,L_{\odot}$. Error bars represent the standard
  deviation.}
\label{fig_simul_sfr}
\end{figure}

We show here the impact of the incompleteness in \luv on the recovered
SFR stellar mass relation. We use the same method as
\citet{Reddy_2012b} as described in Sect. \ref{sec_irx_bias} to create
a mock catalogue. We show in Fig. \ref{fig_simul_sfr} as black circles
the input SFR-Mass relation at $z\sim1.5$ from the mock catalogue we
build. Red circles show the recovered SFR-Mass relation we obtain from
this mock catalogue if we use only objects brighter than $L_{\rm FUV}
>10^{10}\, L_{\odot}$. This shows that there is no impact of UV
incompleteness on the SFR-Mass relation we observe.

\section{Bias on the estimation of dust attenuation relations at $z\sim3,4$}\label{app_simul_irx_mass}
\begin{figure}
\includegraphics[width=\hsize]{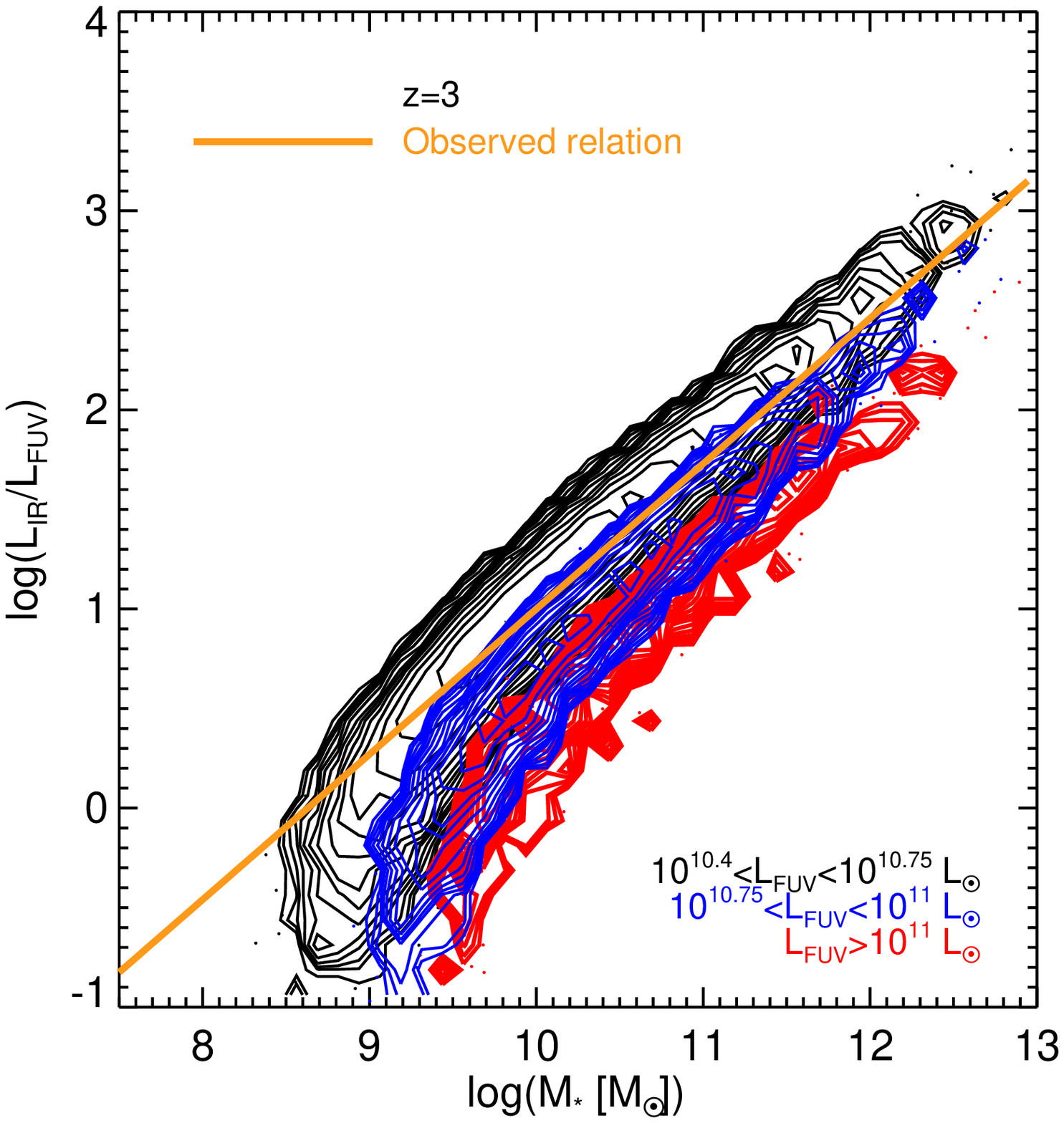}
\caption{Simulated FIR to UV luminosity ratio as a function of stellar
  mass from a mock catalogue (see text) at $z\sim3$. The red contours
  show mock galaxies with $L_{\rm FUV}>10^{11}\,\rm{L}_{\odot}$, blue
  contours galaxies with $10^{10.75}<L_{\rm
    FUV}/\rm{L}_{\odot}<10^{11}$, and black contours galaxies with
  $10^{10.4}<L_{\rm FUV}/\rm{L}_{\odot}<10^{10.75}$.The solid line
  represents the observed relation.}
\label{fig_simul_irx_mass_z3}
\end{figure}

\begin{figure}
\includegraphics[width=\hsize]{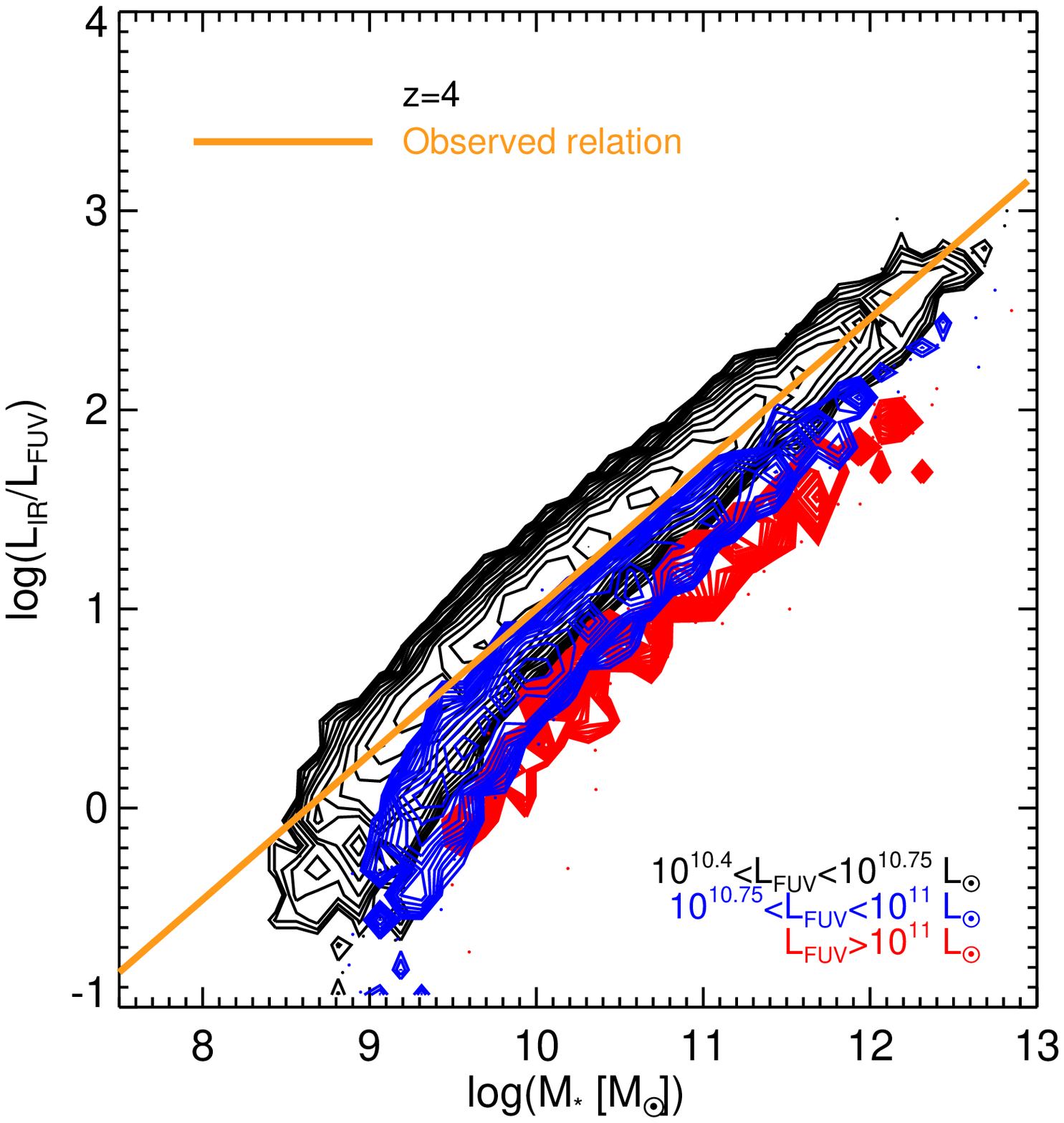}
\caption{Simulated FIR to UV luminosity ratio as a function of stellar
  mass from a mock catalogue (see text) at $z\sim4$. The red contours
  show mock galaxies with $L_{\rm FUV}>10^{11}\,\rm{L}_{\odot}$, blue
  contours galaxies with $10^{10.75}<L_{\rm
    FUV}/\rm{L}_{\odot}<10^{11}$, and black contours galaxies with
  $10^{10.4}<L_{\rm FUV}/\rm{L}_{\odot}<10^{10.75}$.The solid line
  represents the observed relation.}
\label{fig_simul_irx_mass_z4}
\end{figure}

We show here the IR to UV luminosity ratio as a function of stellar
mass from mock catalogues built as described in Sect. \ref{sec_irx_bias}
at $z\sim3$ in Fig. \ref{fig_simul_irx_mass_z3}, and at $z\sim4$ in
Fig. \ref{fig_simul_irx_mass_z4}. Note that here we do not extrapolate
to UV luminosities fainter than the completeness limit of the samples
as we do in Fig. \ref{fig_simul_irx}.
 
\end{document}